\documentclass[aps,prx,twocolumn,superscriptaddress,floatfix]{revtex4-2}

\usepackage{graphicx}   
\usepackage{dcolumn}    
\usepackage{bm}         
\usepackage{amsmath}    
\usepackage{amssymb}
\usepackage{xcolor}
\usepackage{subcaption}
\usepackage{comment}
\usepackage{color}
\usepackage{float}
\usepackage{hyperref}
\usepackage{tabularx}
\usepackage{booktabs}
\usepackage{ragged2e}
\usepackage[justification=justified]{caption}

\begin{document}

\title{Mapping of Fermionic Lattice Models for Ising Solvers}

\author{Lakshya Nagpal}
\email{lakshyan@imsc.res.in}
\affiliation{Institute of Mathematical Sciences CIT Campus, Tharamani 600113}
\affiliation{Pecslab Research}

\author{Aditya Kumar}%
\affiliation{Indian Institute of Science Education and Research Pune  }%

\author{S. R. Hassan}
\affiliation{Institute of Mathematical Sciences CIT Campus, Tharamani 600113}

\date{\today}

\begin{abstract}
We present an end-to-end, symmetry-aware pipeline that converts interacting fermionic and quantum-spin models into annealer-ready QUBOs while preserving low-energy physics. The workflow combines Bravyi–Kitaev encoding, exact $\mathbb{Z}_2$ symmetry tapering, Xia–Bian–Kais (XBK) diagonalization to a Z-only form, and $k\!\to\!2$ quadratization, with ground energies recovered via a Dinkelbach fixed-point over the resulting Ising objective. We validate the approach across a complexity ladder: (i) a frustrated 2D Ising model run on a D-Wave Advantage QPU reproduces the known ferromagnet–stripe transition; (ii) finite-temperature checks on 1D Ising recover standard finite-size trends; (iii) a genuinely quantum spin target (XXZ) matches exact diagonalization (ED) on small chains; and (iv) interacting fermions (t–V) in 1D (rings $L=2\!-\!8$) show ED-level energies and the expected kink near $V/t\approx2$, with a 2D $2\times2$ cluster tracking ED slopes up to a uniform offset. A replication-factor study quantifies the accuracy–overhead trade-off, with $\mathcal{O}$-of-magnitude error reduction by  and diminishing returns beyond $r\!\approx\!N_q$ Except for the classical Ising benchmark and Molecular benchmarks, experiments use D-Wave’s public DIMOD and Neal simulators; a molecular benzene case in the appendix illustrates portability beyond lattices. The results establish a practical pathway for mapping quantum matter to current annealers, with clear knobs for fidelity, resources, and embedding.

\end{abstract}

\maketitle

\section{Introduction}

Interacting quantum lattices—fermions, spins, and bosons—exhibit superconductivity, quantum magnetism, correlated metals, and topological order. Predictive simulation is hard because Hilbert spaces grow exponentially and quantum statistics impose long-range constraints.\cite{schollwock_density-matrix_2011,orus_practical_2014}

Classical solvers each cover part of this space: exact diagonalization (ED) is definitive but small-scale; quantum Monte Carlo (QMC) reaches larger sizes but suffers from the sign problem in frustrated/fermionic regimes;\cite{troyer_computational_2005} tensor networks, especially DMRG, excel in 1D and some quasi-2D cases but degrade with entanglement and topology.\cite{white_density_1992,schollwock_density-matrix_2011,orus_practical_2014}

Quantum hardware offers a complementary route. Gate-based processors can implement the full operator algebra but are currently limited by qubit counts, fidelities, and depth.\cite{preskill_quantum_2018} Quantum annealers already provide thousands of qubits with persistent couplings and directly minimize Ising/QUBO cost functions,\cite{lucas_ising_2014,boothby_next-generation_2020} provided physical Hamiltonians—typically noncommuting and sometimes multi-body—are translated to a diagonal, two-local Ising form without losing essential physics.

\paragraph*{From fermions to qubits: where BK and XBK fit.}
Fermion-to-qubit encodings map second-quantized operators to Pauli strings. Jordan–Wigner (JW) and Bravyi–Kitaev (BK) produce a Pauli (spin/qubit) Hamiltonian usable on gate-based devices (Trotterization or VQE),\cite{jordan_uber_1928,bravyi_fermionic_2002,seeley_bravyi-kitaev_2012,lloyd_universal_1996,peruzzo_variational_2014} but present hardware remains resource-limited.\cite{preskill_quantum_2018} In higher-dimensional (higher-D) settings, JW parity strings grow long; BK distributes parity/occupation to reduce typical operator weight (from $\mathcal{O}(L)$ toward $\mathcal{O}(\log L)$), especially with lattice \emph{linearizations} (e.g., Hilbert-curve orderings, a class of space-filling curves) that keep geometric neighbors adjacent in index space.\cite{havlicek_operator_2017,steudtner_quantum_2019} For annealers the Pauli form is not yet “ready,” because $X$/$Y$ terms are nondiagonal. The Xia–Bian–Kais (XBK) method embeds the problem in a replicated register and rewrites all $X$/$Y$ structure as $Z$-only couplings—with sector bookkeeping—yielding a diagonal Ising Hamiltonian at the cost of replication.\cite{xia_electronic_2021,xia_electronic_2017}

\paragraph*{What is new here.}
We assemble a deterministic, symmetry-aware pipeline tailored to lattice models and current annealing hardware:
(i) \textbf{BK} (or direct spin form) for a locality-friendly Pauli Hamiltonian;\cite{bravyi_fermionic_2002,seeley_bravyi-kitaev_2012}
(ii) \textbf{exact $\boldsymbol{\mathbb{Z}_2}$ tapering} \emph{before} diagonalization to remove qubits without approximation and fix the symmetry sector;\cite{bravyi_tapering_2017}
(iii) \textbf{XBK} to obtain a $Z$-only Hamiltonian with sector bookkeeping;\cite{xia_electronic_2021}
(iv) \(k\!\to\!\)two reduction with correctness-preserving penalties to form a QUBO;\cite{rosenberg_reduction_1975,ishikawa_transformation_2011}
(v) sector energies via a discrete Rayleigh quotient solved by \textbf{Dinkelbach iteration} for monotone convergence diagnostics.\cite{Dinkelbach_1967}
We use hardware-aware choices (lattice ordering, penalty scales, and Pegasus-compatible minor-embedding/graph shaping).\cite{boothby_next-generation_2020}

\begin{figure*}[t]
    \centering
    \includegraphics[width=1.0\textwidth]{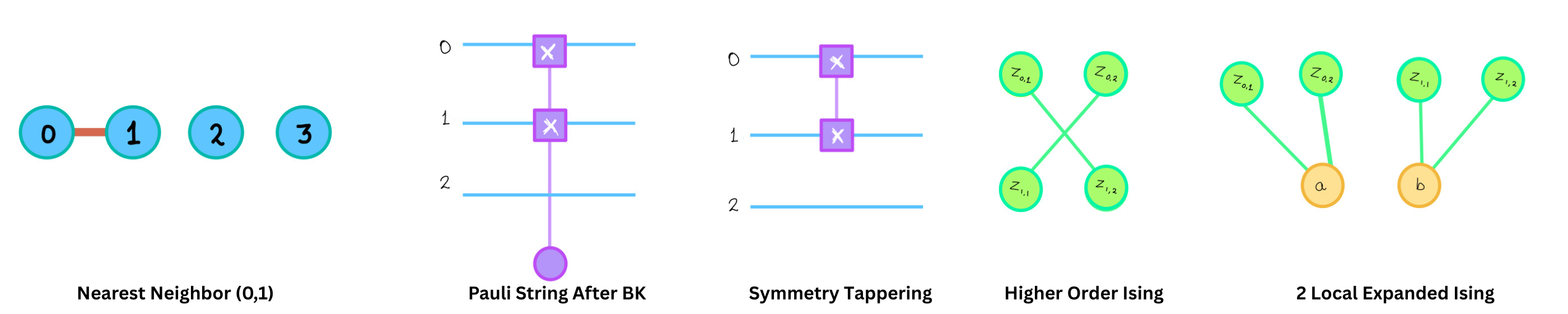}
    \caption{\justifying Deterministic, symmetry-aware pipeline from a second-quantized (or spin) Hamiltonian to a \textbf{2-local} QUBO: Bravyi–Kitaev (BK) mapping, $\mathbb{Z}_2$ symmetry tapering, Xia–Bian–Kais (XBK) diagonalization, $k\!\to\!2$ reduction, and final \textbf{2-local} QUBO assembly.}
    \label{fig:pipeline}
\end{figure*}

\paragraph*{Positioning relative to alternatives.}
JW+XBK without tapering inflates nonlocal strings and replicas on higher-D lattices;\cite{seeley_bravyi-kitaev_2012,xia_electronic_2017} parity mappings with tapering remove global symmetries but typically yield longer strings than BK under standard 2D orderings. Direct quadratization of nondiagonal Pauli terms (skipping XBK) forces large penalties and risks spurious minima.\cite{choi_reducing_2011,choi_minor-embedding_2011,lucas_ising_2014} Gate-based routes avoid Ising-only reduction but currently face depth/fidelity limits.\cite{preskill_quantum_2018} Our BK\,$\to$\,tapering\,$\to$\,XBK\,$\to$\,\(k\!\to\!\)two pipeline balances fidelity, variable growth, and embeddability for extended lattices.

We benchmark three models: a frustrated classical Ising system, the spin-$\tfrac{1}{2}$ XXZ chain, and the spinless fermionic $t$–$V$ model. In each case the resulting two-local Ising Hamiltonian is \emph{annealer-ready}. The frustrated classical Ising benchmark was executed on a D\mbox{-}Wave Advantage QPU to validate end-to-end embedding and readout, whereas the XXZ and spinless $t$–$V$ studies used the D\mbox{-}Wave Ocean DIMOD simulators (no sustained QPU access for these cases) these were further compared with neal (simulated annealing) packages to verify QUBO acceptance and embedding. Reported sizes for XXZ and $t$–$V$ therefore reflect simulator limits rather than QPU capacity. Although our focus is on lattices, the pipeline is portable; we also apply it to a small molecular Hamiltonian (benzene) in Appendix~E, and summarize a lightweight GUI in Appendix~F.

\emph{Roadmap.} Section~II defines models and observables. Section~III details the pipeline (BK, higher-D ordering, exact $\mathbb{Z}_2$ tapering, XBK, \(k\!\to\!\)two reduction, Dinkelbach solver). Section~IV quantifies resource growth (tapered qubits, replication $r$, auxiliaries, minor-embedding). Section~V presents benchmarks; Section~VI contrasts asymptotic costs with DMRG; Section~VII outlines limitations and extensions.

\section{Model Selection}

We consider two complementary model families that together span classical and fully quantum regimes while exercising the benchmarks used in this work: a \emph{fermionic} lattice Hamiltonian and a \emph{spin} (qubit) Hamiltonian. The former stresses fermion-to-qubit encodings and symmetry structure; the latter isolates noncommuting spin dynamics without fermionic statistics.

This section fixes the model families and parameter conventions used throughout. The complete mapping pipeline—fermion-to-qubit encoding, exact $\mathbb{Z}_2$ tapering, XBK diagonalization, and $k\!\to\!2$ reduction to QUBO—is presented in Sec.~III.

\subsection{Fermionic class}

Our fermionic baseline is a spinless tight-binding model with density--density interactions,
\small{
\begin{equation}
\label{eq:H_fermion}
H_{\mathrm{f}}
= -\sum_{\langle i,j\rangle} t_{ij}\,\big(c_i^\dagger c_j + c_j^\dagger c_i\big)
  + \sum_{\langle i,j\rangle} V_{ij}\,\Big(n_i-\tfrac{1}{2}\Big)\Big(n_j-\tfrac{1}{2}\Big),
\qquad n_i=c_i^\dagger c_i.
\end{equation}
}
Nearest-neighbor couplings $t_{ij}=t$ and $V_{ij}=V$ recover the standard \emph{$t$--$V$ model}. In 1D at half filling (with the standard Jordan--Wigner convention), the JW transform maps \eqref{eq:H_fermion} to an XXZ spin chain with couplings $J_\perp=-2t$ and $J_z=V$ (up to constants), i.e.\ anisotropy $\Delta=J_z/J_\perp=V/(-2t)$.\cite{jordan_uber_1928} In higher dimensions, JW parity strings grow long; we therefore adopt the Bravyi--Kitaev (BK) mapping to reduce operator weight and improve locality before later stages.\cite{bravyi_fermionic_2002,seeley_bravyi-kitaev_2012,havlicek_operator_2017,steudtner_quantum_2019}

\subsection{Spin class}

A single anisotropic nearest-neighbor spin Hamiltonian with optional fields unifies the cases of interest,
\begin{equation}
\label{eq:H_spin}
H_{\mathrm{s}}
= \sum_{\langle i,j\rangle}\!\big(J_x S_i^x S_j^x + J_y S_i^y S_j^y + J_z S_i^z S_j^z\big)
  - \sum_i \big(h_x S_i^x + h_z S_i^z\big),
\end{equation}
with $S_i^\alpha$ the spin-$\tfrac{1}{2}$ operators (we take $S_i^\alpha=\tfrac{1}{2}\sigma_i^\alpha$). Specializations used below include:
(i) \textbf{Ising} (classical/QUBO baseline): $J_z\!\neq\!0$, $J_x=J_y=0$, $h_x=0$ with optional $h_z$;
(ii) \textbf{Transverse-field Ising (TFIM)}: $J_z\!\neq\!0$, $J_x=J_y=0$, $h_x\!\neq\!0$ (noncommuting);
(iii) \textbf{XXZ}: $J_x=J_y\equiv J_\perp$, $\Delta=J_z/J_\perp$ (the \textbf{Heisenberg} point is $\Delta=1$).

Section~III details the mapping pipeline and the hardware-aware choices (lattice linearizations, symmetry tapering, XBK parameters, and $k\!\to\!2$ reduction) used to obtain annealer-ready QUBO instances for the models defined above.

\section{Methodology: Mapping to an Ising-Compatible Hamiltonian}
\label{sec:method}

Executing fermionic or spin-lattice models on a quantum annealer requires transforming the original Hamiltonian into a two-local, \emph{diagonal} Ising form compatible with the device’s coupling topology. We follow a five-stage pipeline:

\begin{enumerate}
    \item \textbf{Bravyi--Kitaev (BK) encoding:} Map the fermionic Hamiltonian to an exact Pauli-operator form with short typical strings.\cite{bravyi_fermionic_2002,seeley_bravyi-kitaev_2012}
    \item \boldmath$\mathbb{Z}_2$ \textbf{symmetry tapering:} Identify commuting symmetries and project onto fixed sectors, eliminating redundant qubits.\cite{bravyi_tapering_2017}
    \item \textbf{Xia--Bian--Kais (XBK) embedding:} Replicate qubits to replace all $X$ and $Y$ operators with $Z$-only couplings via sector-dependent signs.\cite{xia_electronic_2021}
    \item \textbf{Quadratization:} Reduce any $k$-local ($k>2$) Ising terms to two-local form using auxiliary binary variables and penalty constraints.\cite{rosenberg_reduction_1975,choi_reducing_2011,ishikawa_transformation_2011,boros_pseudo-boolean_2002,glover_tutorial_2018}
    \item \textbf{Classical/QUBO solving:} Minimize the resulting QUBO; recover ground-state energies via a Dinkelbach-style fixed-point refinement of a discrete Rayleigh quotient.\cite{Dinkelbach_1967}
\end{enumerate}

This chain preserves the essential spectrum within the chosen symmetry sector while yielding an annealer-ready formulation directly embeddable on present-day hardware.\cite{boothby_next-generation_2020,lucas_ising_2014,hauke_perspectives_2020}

\subsection{Bravyi--Kitaev Mapping (BK)}
\label{sec:bk}

Fermionic Hamiltonians in second quantization must be recast into qubit operators to run on quantum hardware. The BK transformation offers an efficient compromise between Jordan--Wigner (local updates but long parity strings in higher-D) and parity encodings (short parity, heavy updates), reducing many operator weights from $\mathcal{O}(L)$ to $\mathcal{O}(\log L)$ by distributing occupation/parity information across a Fenwick tree (binary indexed tree).\cite{bravyi_fermionic_2002,seeley_bravyi-kitaev_2012}

\begin{figure}[t]
    \centering
    \includegraphics[scale=0.3]{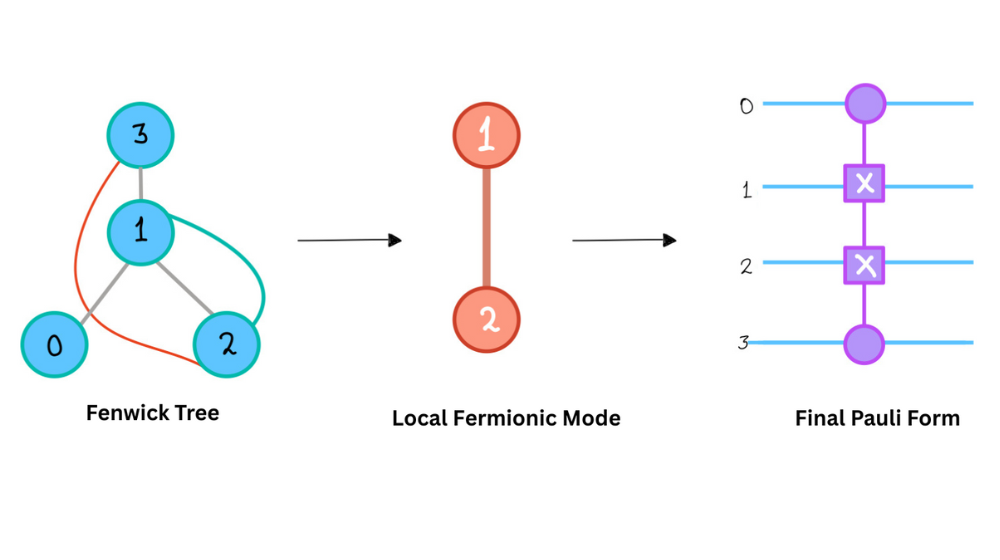}
    \caption{\justifying\textbf{Fenwick-tree view of Bravyi--Kitaev indexing (example $L\!=\!4$).}
    Blue arrows trace \emph{parity propagation} (build $P(j)$ by repeatedly subtracting $\mathrm{lowbit}$);
    red arrows trace \emph{update propagation} (build $U(j)$ by repeatedly adding $\mathrm{lowbit}$).
    This makes the composition of the BK sets $U(j)$, $P(j)$, $F(j)$, and $R(j)$ explicit.}
    \label{fig:fenwick_tree}
\end{figure}

\paragraph{BK sets and Fenwick-tree picture.}
For mode $j$ (0-based), four sets determine the mapping:
\emph{Update} $U(j)$, \emph{Parity} $P(j)$, \emph{Flip} $F(j)\!\subseteq\!P(j)$, and \emph{Remainder} $R(j)=P(j)\setminus F(j)$.
They are generated by the \emph{lowbit} primitive $\mathrm{lowbit}(m)=m\,\&\,(-m)$ on $m=j+1$ using Fenwick-tree forward (updates) and backward (parity) traversals; Fig.~\ref{fig:fenwick_tree} visualizes this propagation.

\begin{figure}[t]
    \centering
    \includegraphics[width=0.82\linewidth]{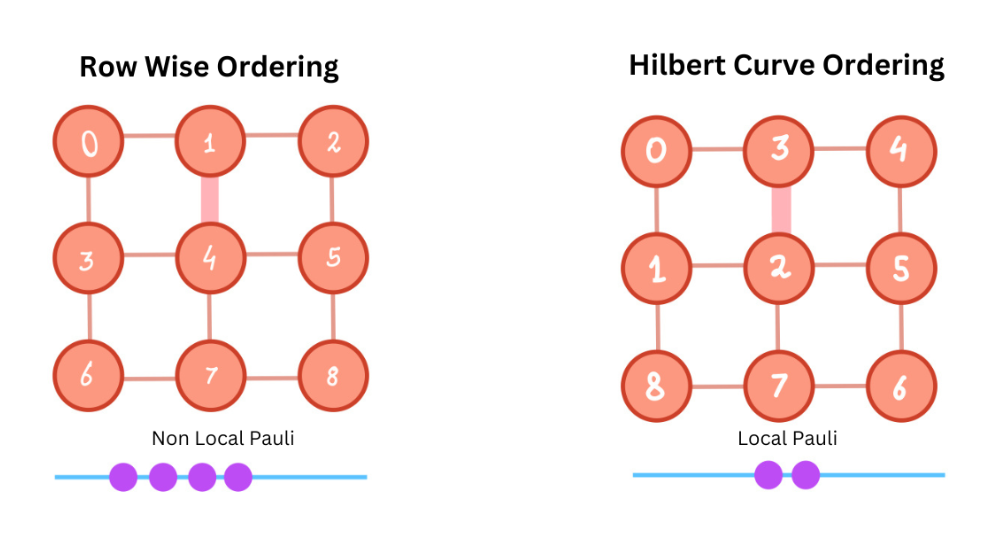}
    \caption{\justifying \textbf{Index ordering.} In a $3\times 3$ lattice, a row-wise linearization pushes vertical neighbors far apart in index space, yielding long parity strings (left). A Hilbert-curve ordering keeps geometric neighbors close, shortening strings and improving locality (right).\cite{havlicek_operator_2017}}
    \label{fig:ordering_locality}
\end{figure}

\paragraph{Ordering in higher dimensions.}
Although the BK sets depend only on $L$ and $j$, the \emph{ordering} of modes is induced by the lattice geometry.
Row-wise or column-wise enumerations preserve nearest neighbors in only one direction, whereas \emph{space-filling curves} (e.g., the Hilbert curve) keep horizontal and vertical neighbors close in index space, shortening parity strings and improving hardware compatibility.\cite{havlicek_operator_2017,steudtner_quantum_2019}
Figure~\ref{fig:ordering_locality} illustrates the reduction in Pauli-string nonlocality for a $3\times 3$ lattice.

\paragraph{Operator form.}
Let $\rho(j)=P(j)$ for even $j$ and $\rho(j)=R(j)$ for odd $j$.
The BK images of $c_j^\dagger$ and $c_j$ are
\begin{align}
    c_j^\dagger &= \tfrac12\!\left[
        \Big(\!\bigotimes_{k\in U(j)}\!X_k\!\Big) X_j \Big(\!\bigotimes_{k\in P(j)}\!Z_k\!\Big)
        \;-\; i\,\Big(\!\bigotimes_{k\in U(j)}\!X_k\!\Big) Y_j \Big(\!\bigotimes_{k\in \rho(j)}\!Z_k\!\Big)
    \right], \label{eq:bk_cre}\\
    c_j &= \tfrac12\!\left[
        \Big(\!\bigotimes_{k\in U(j)}\!X_k\!\Big) X_j \Big(\!\bigotimes_{k\in P(j)}\!Z_k\!\Big)
        \;+\; i\,\Big(\!\bigotimes_{k\in U(j)}\!X_k\!\Big) Y_j \Big(\!\bigotimes_{k\in \rho(j)}\!Z_k\!\Big)
    \right]. \label{eq:bk_ann}
\end{align}

A worked $L{=}4$ example (including explicit $U$, $P$, $F$, $R$ sets and the qubit Hamiltonian) is provided in Appendix~\ref{ap:L4ex_bk}.

\subsection{Exact Qubit Reduction via $\mathbb{Z}_2$ Symmetry Tapering}
\label{sec:tapering}

\begin{figure}[h!]
    \centering
    \includegraphics[scale=0.3]{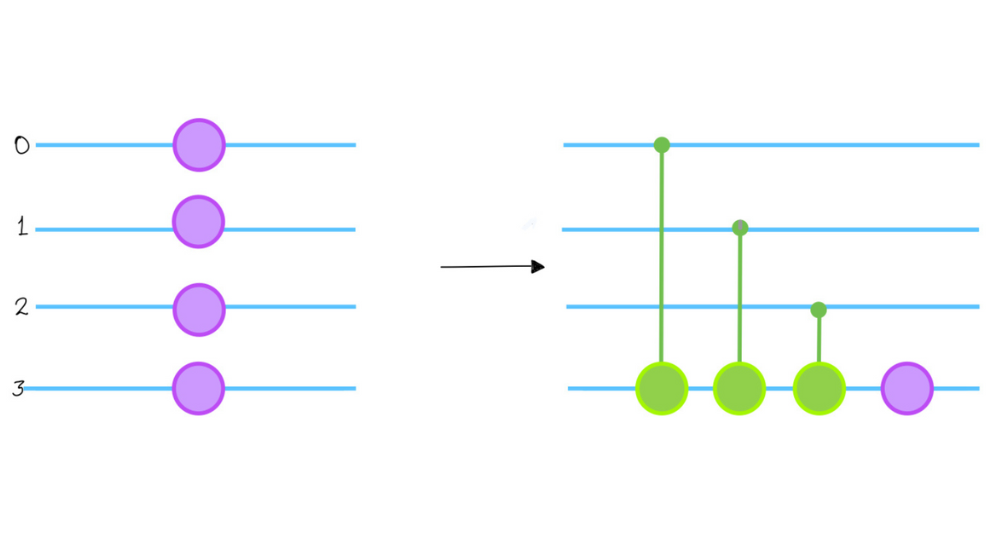}
    \caption{\justifying \textbf{Symmetry tapering workflow.}
    \textbf{(a)} Encode Hamiltonian Pauli terms as symplectic rows of $E$; commuting $\mathbb{Z}_2$ symmetries are the nullspace of $EJ$.
    \textbf{(b)} Map a symmetry vector back to a Pauli $\tau$ and use local Clifford operations (e.g., CNOT fan-in) to reduce $\tau$ to a single $Z_q$.
    \textbf{(c)} Fix the symmetry sector ($Z_q\!=\!\lambda$) to \emph{remove} qubit $q$ from the Hamiltonian exactly. Applying tapering \emph{before} XBK exposes symmetries that would otherwise be hidden once all operators are diagonalized.}
    \label{fig:tapering}
\end{figure}

Post-BK, many models retain discrete symmetries that are manifest in Pauli form. Tapering eliminates qubits \emph{exactly} whenever the Hamiltonian commutes with a Pauli operator $\tau$,
$[\tau,H]=0$.\cite{bravyi_tapering_2017}
Writing each Pauli string $P$ as a symplectic vector
$\mathbf{v}=(x_1,\dots,x_n\mid z_1,\dots,z_n)\in\mathbb{F}_2^{2n}$ with $I\!\to(0,0)$,
$X\!\to(1,0)$, $Z\!\to(0,1)$, $Y\!\to(1,1)$, two strings commute if and only if $\mathbf{v}_a J \mathbf{v}_b^T=0 \pmod{2}$, where $J=\begin{bmatrix}0&I\\ I&0\end{bmatrix}$.\cite{gottesman_stabilizer_1997,aaronson_improved_2005}
Stacking the $m$ Hamiltonian terms into $E\in\mathbb{F}_2^{m\times 2n}$, all $\mathbb{Z}_2$ symmetries are obtained from the nullspace of $EJ$:
\[
EJ\,\mathbf{u}^T=\mathbf{0}\pmod{2}.
\]
Each $\mathbf{u}$ maps back to a Pauli $\tau$; local Clifford gates (single-qubit $H,S$ and CNOTs) are then used to convert $\tau$ to a single $Z_q$.\cite{nielsen_quantum_2010} Fixing the sector ($Z_q\!\to\!\lambda$, $\lambda=\pm1$) removes qubit $q$ exactly. Figure~\ref{fig:tapering} summarizes this workflow.

\paragraph{Example (global $Z$-parity, $L=4$).}
For the BK form of the $L{=}4$ spinless $t$--$V$ chain (see Appendix~\ref{ap:L4ex_bk}), encoding representative terms into $\mathbb{F}_2^{8}$ yields a nullspace generator
$\mathbf{u}_{\rm parity}=(0,0,0,0\mid 1,1,1,1)$, corresponding to $\tau=Z_0Z_1Z_2Z_3$.
A short Clifford sequence
$\mathrm{CNOT}_{0\to 3}\,\mathrm{CNOT}_{1\to 3}\,\mathrm{CNOT}_{2\to 3}$
maps $\tau\mapsto Z_3$, after which setting $Z_3\!\to\!+1$ removes qubit~3 exactly.
Applying tapering \emph{before} XBK is crucial: with $X/Y$ present, symmetries are visible; after XBK the Hamiltonian is diagonal and no new commuting structure emerges.\cite{bravyi_tapering_2017,xia_electronic_2017}
A worked example of qubit tapering for the $L{=}4$ spinless $t$--$V$ model is provided in Appendix~\ref{ap:tapering_L4}.

\subsection{From Qubit Tapering to XBK Diagonalization}
\label{sec:xbk}

After tapering, the Hamiltonian still contains $X$ and $Y$ operators. Quantum annealers natively implement $Z$-only Ising Hamiltonians, so we adopt XBK to embed the model into a replica-augmented register where every Pauli factor is rewritten as a $Z$-only expression across two copies.\cite{xia_electronic_2017,xia_electronic_2021}
Let $m$ be the tapered-qubit count and $r\!\ge\!2$ the replication factor. Copy $j{=}1$ is the \emph{reference}, $j\!>\!1$ are \emph{replicas}. For each sector $p$, a sign table $S_p(i,j)\in\{\pm1\}$ encodes relative phases; by convention $S_p(i,1)\equiv +1$.

For a fixed pair $(j,k)$ with $j{=}1$ and $k\!>\!1$, single-qubit Paulis map as
\begin{equation}
\begin{aligned}
\sigma_x^{(i)} &\longrightarrow S_p(i,j) S_p(i,k) \, \frac{1 - \sigma_z^{(i_j)} \sigma_z^{(i_k)}}{2}, \\
\sigma_y^{(i)} &\longrightarrow i \, S_p(i,j) S_p(i,k) \, \frac{\sigma_z^{(i_k)} - \sigma_z^{(i_j)}}{2}, \\
\sigma_z^{(i)} &\longrightarrow S_p(i,j) S_p(i,k) \, \frac{\sigma_z^{(i_j)} + \sigma_z^{(i_k)}}{2}, \\
I^{(i)}       &\longrightarrow \frac{1 + \sigma_z^{(i_j)} \sigma_z^{(i_k)}}{2},
\end{aligned}
\label{eq:xbk_mapping}
\end{equation}
so each non-diagonal factor becomes a $Z$-bilinear across \emph{copies}. Products of Paulis are mapped multiplicatively.

\begin{figure}[tb]
    \centering
    \includegraphics[scale=0.35]{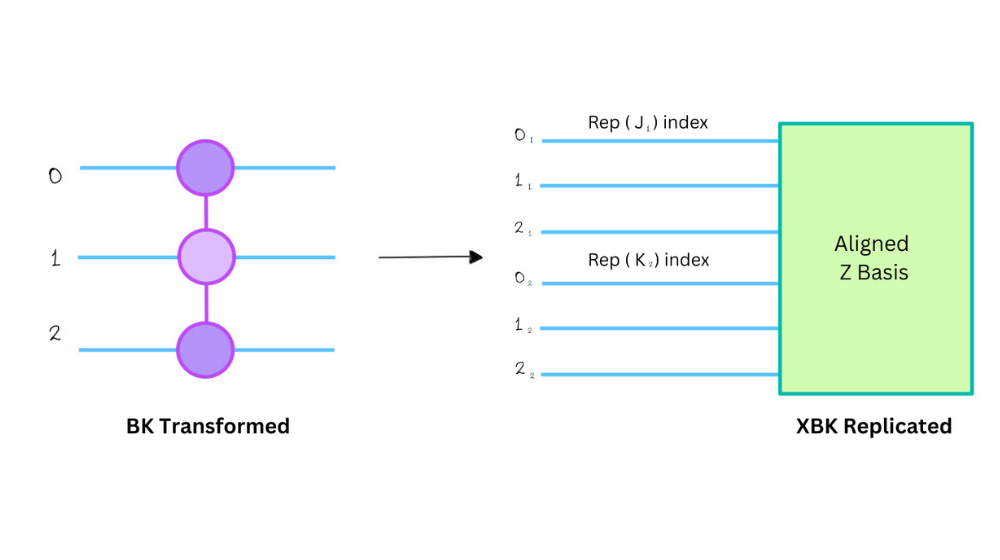}
    \caption{\justifying \textbf{XBK pairwise-copy mapping.} Each non-diagonal single-qubit Pauli ($X$ or $Y$) is rewritten as a bilinear in $Z$ across a reference copy ($j{=}1$) and a replica ($k{>}1$), controlled by sector-dependent signs $S_p(i,j)S_p(i,k)$. Multi-qubit Pauli strings map multiplicatively, yielding a diagonal ($Z$-only) Ising Hamiltonian on $rm$ qubits.}
    \label{fig:xbk_schematic}
\end{figure}

\paragraph{XBK example.}
For $r=2$ and a BK term $X_2 Y_0 Z_1$, choose sector signs with
$S_p(0,2)=-1$, $S_p(1,2)=+1$, $S_p(2,2)=-1$.
Equation~\eqref{eq:xbk_mapping} gives
\[
X_2 \ \to\ (+1)\,\frac{1-Z^{(2_1)}Z^{(2_2)}}{2},\quad
Y_0 \ \to\ i(-1)\,\frac{Z^{(0_2)}-Z^{(0_1)}}{2},\quad
Z_1 \ \to\ (+1)\,\frac{Z^{(1_1)}+Z^{(1_2)}}{2},
\]
so the image is $Z$-only on the $2m$-qubit replicated space.

Because replication duplicates logical basis states, a \emph{counting} functional $C_p(\vec{s})>0$ normalizes multiplicities.\cite{xia_electronic_2021}
For sector $p$ we minimize the discrete Rayleigh quotient
\begin{equation}
E_{0,p}=\min_{\vec{s}\in\{\pm1\}^{rm}} \frac{H'_p(\vec{s})}{C_p(\vec{s})}, 
\qquad
E_0=\min_p E_{0,p},
\label{eq:discreteRQ}
\end{equation}
or, equivalently, solve $\phi_p(\lambda)=\min_{\vec{s}}\!\big[H'_p(\vec{s})-\lambda\,C_p(\vec{s})\big]=0$ by Dinkelbach iteration.\cite{Dinkelbach_1967}
XBK yields an Ising-only form but introduces $k$-local ($k>2$) $Z$ products, addressed next via quadratization.

A detailed derivation and implementation notes for XBK are provided in Appendix~\ref{ap:Detail_xbk}.

\subsection{Reduction of High-Order Ising Terms to Quadratic Form}
\label{sec:quadratization}

XBK images generally contain higher-order $Z$ products. Because current annealers natively realize only two-local couplings, we reduce $k$-local terms to quadratic form via standard pseudo-Boolean reductions.\cite{rosenberg_reduction_1975,choi_minor-embedding_2011,ishikawa_transformation_2011,boros_pseudo-boolean_2002,glover_tutorial_2018}

\paragraph*{Variable domain and notation.}
After XBK, we work with Ising spins $s_\ell\!\in\!\{\pm1\}$. For quadratization we introduce binary variables $x_\ell\!\in\!\{0,1\}$ via
\[
x_\ell=\tfrac{1-s_\ell}{2}
\quad\Longleftrightarrow\quad
s_\ell=1-2x_\ell,
\]
so that products of $Z$ operators map to monomials in the $x$’s. (Conversion back to Ising couplings for embedding is straightforward: $x_i x_j=\tfrac{1}{4}(1-s_i-s_j+s_i s_j)$.)

\paragraph*{Auxiliary-product gadgets.}
To reduce a cubic monomial $x_i x_j x_k$, introduce an auxiliary $a=x_i x_j$ and enforce the identity with a quadratic penalty. A common Rosenberg/Choi form is
\begin{equation}
\label{eq:penalty}
P(x_i,x_j,a)=M\big(x_i x_j - 2 a x_i - 2 a x_j + 3 a\big),
\end{equation}
with $M>0$ chosen large enough that any violation of $a=x_i x_j$ is energetically disfavored. Then
\[
x_i x_j x_k \ \Longrightarrow\ a x_k + P(x_i,x_j,a),
\]
and the same construction is applied recursively (e.g., define $b=a x_c$ to break a remaining 3-local factor), until only pairwise terms remain. When expressed directly in Ising form, \eqref{eq:penalty} yields linear fields and pairwise couplers only.

\paragraph*{Choosing the penalty weight $M$.}
Let $c_\alpha$ denote the coefficients of terms that involve the auxiliary variable(s) created by a reduction step. A sufficient, implementation-friendly choice is
\[
M \ \ge\ 1+\sum_\alpha |c_\alpha|,
\]
which prevents any spurious minimum caused by violating the constraint from beating the best feasible assignment. Tighter, instance-specific $M$’s (e.g., per gadget) improve scaling and are used in our code.

\paragraph*{Special cases and improvements.}
For submodular terms (certain negative-coefficient structures), Ishikawa’s construction can reduce degree without auxiliaries; otherwise, Rosenberg/Choi gadgets are near-minimal in auxiliary count.\cite{rosenberg_reduction_1975,ishikawa_transformation_2011,boros_pseudo-boolean_2002} We follow the minimal-auxiliary chain gadgets for general $k$-local terms.

\begin{figure}[t]
\centering
\includegraphics[scale=0.3]{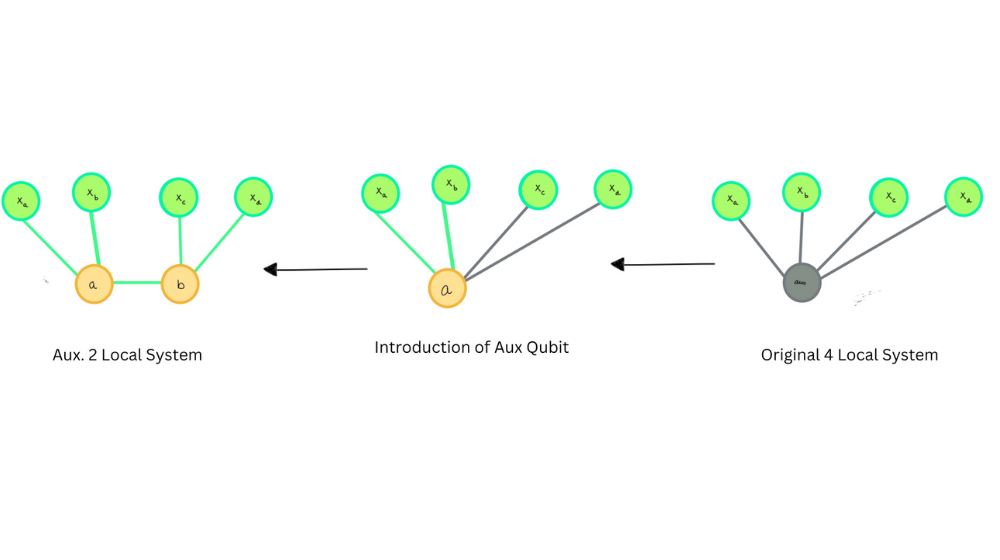}
\caption{\justifying \textbf{Quadratization schematic (example $4$-local).} 
A $4$-local term $x_a x_b x_c x_d$ (right) is reduced by introducing $a=x_a x_b$ (center) and then $b=a x_c$ (left), with penalties (dashed) enforcing each definition. The final graph contains only pairwise (2-local) interactions, i.e., a valid QUBO instance.}
\label{fig:k_local_reduction}
\end{figure}

This procedure guarantees a QUBO form while preserving the ground-state manifold (for $M$ above the stated threshold). The increase in variables from auxiliaries is accounted for in Sec.~\ref{sec:overhead}.
\vspace{0.5em}

\subsection{From Qubit Hamiltonian to Classical Optimization}
\label{sec:classical}

For each symmetry sector $p$, tapering+XBK yields a diagonal Ising Hamiltonian $H'_p(\vec{s})$ on $rm$ spins and a positive counting functional $C_p(\vec{s})$ that normalizes replica multiplicities.\cite{xia_electronic_2021}
We recover the ground-state energy via the discrete Rayleigh quotient
\[
E_{0,p}=\min_{\vec{s}\in\{\pm1\}^{rm}} \frac{H'_p(\vec{s})}{C_p(\vec{s})},
\qquad
E_0=\min_p E_{0,p},
\]
or, equivalently, by solving the Dinkelbach transform
\[
\phi_p(\lambda)=\min_{\vec{s}} \big[\,H'_p(\vec{s})-\lambda\,C_p(\vec{s})\,\big]=0.
\]
The root $\lambda=E_{0,p}$ is unique and the (exact) fixed-point iteration is monotone:\cite{Dinkelbach_1967}
\begin{enumerate}
\item Initialize $\lambda_0$ (e.g., from a feasible ratio or a coefficient bound).
\item Approximately minimize $H'_p(\vec{s})-\lambda_k C_p(\vec{s})$ (we use Dwave DIMOD Solver/ parallel tempering) to obtain $\vec{s}_k$.
\item Update $\displaystyle \lambda_{k+1}=\frac{H'_p(\vec{s}_k)}{C_p(\vec{s}_k)}$.
\item Stop when $|\phi_p(\lambda_k)|<\varepsilon$; return $E_{0,p}\approx \lambda_k$.
\end{enumerate}
We repeat across sectors and take $E_0=\min_p E_{0,p}$. The same inner minimization can be run on a QPU by submitting a sequence of QUBOs parameterized by $\lambda_k$; in this work we use classical SA/PT for reproducibility while keeping the objective faithful to the XBK spectrum through the ratio structure $H'_p/C_p$.\cite{xia_electronic_2017}

\section{Results and Discussion}
\label{sec:results}

Our validation strategy follows a complexity hierarchy: (1) purely classical spin models with well-established behavior; (2) genuinely quantum spin systems with noncommuting terms; and (3) interacting fermionic Hamiltonians. This progression benchmarks each stage of the mapping–embedding–optimization pipeline on problems of increasing physical richness and helps isolate potential systematic biases before tackling the full fermionic targets. As emphasized in the Introduction, our pipeline is designed for D-Wave annealers and yields annealer-ready QUBO instances; in this study, however, only the frustrated classical Ising benchmark was executed on a D-Wave Advantage QPU. The XXZ and spinless $t$–$V$ studies were run using D-Wave’s publicly available Ocean simulators, so the reported system sizes for those models reflect simulator and runtime limits rather than QPU capacity. Finally, to demonstrate portability beyond lattices, we also apply the same pipeline to a molecular Hamiltonian (benzene); the end-to-end mapping and results are summarized in Appendix~E.

\subsection{Benchmark I: Frustrated 2D Ising Model at $T=0$}
\label{sec:bench_ising2d}

The first benchmark targets a classical $T=0$ ground-state problem: the square-lattice frustrated Ising model.
Competing ferromagnetic and antiferromagnetic couplings generate a highly degenerate manifold in the intermediate frustration regime, making this a stringent test of optimization quality.

Previous works have implemented this model directly on D-Wave processors for large lattices, including the $20\times20$ case~\cite{kadowaki_quantum_1998,lucas_ising_2014}.
Here, we use it to assess how well the QPU (quantum processing unit) resolves the known phase boundary when the problem is provided in full QUBO form through our pipeline.
We adopt the same $20\times20$ lattice size as Ref.~\cite{kadowaki_quantum_1998} to enable a direct comparison.
The goal is not new physics, but a baseline annealer benchmark on a well-characterized classical instance before moving to quantum systems.

\begin{figure}[htbp]
\centering
\includegraphics[width=\columnwidth]{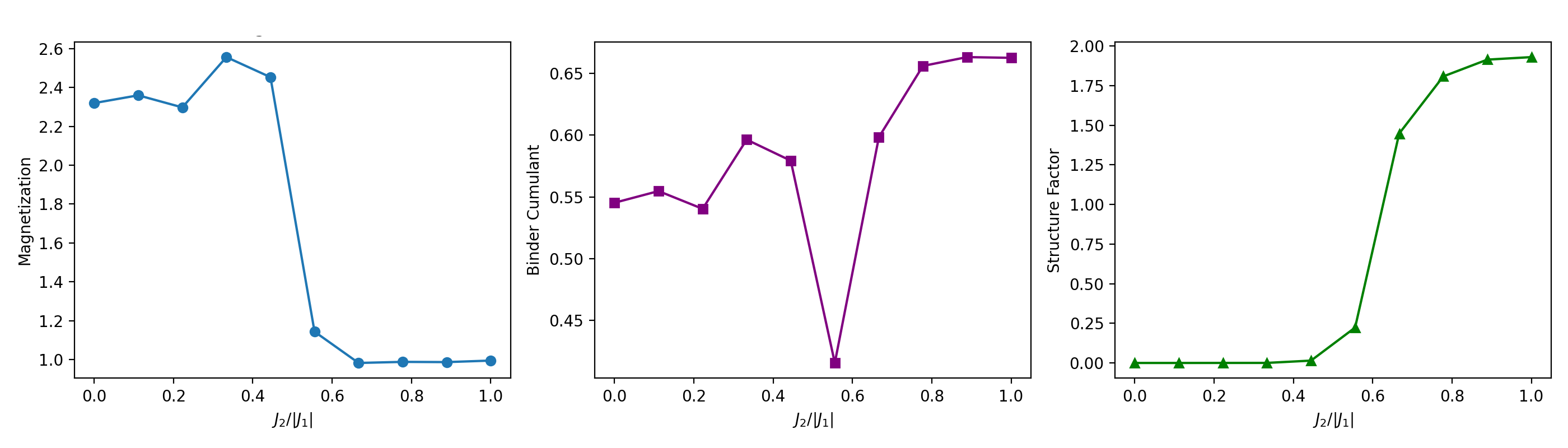}
\caption{\justifying Observables for the frustrated 2D Ising model on a $20 \times 20$ lattice from the D-Wave QPU. Left: Magnetization per spin $M$. Center: Binder cumulant $U$. Right: Structure factor $S(\mathbf{q})$ at the ordering wavevector. The simultaneous drop in $M$ and $U$ and rise in $S(\mathbf{q})$ near $R \approx 0.5$ marks the transition from ferromagnetic to striped order.}
\label{fig:qpu_results}
\caption{\justifying QPU analysis for the $20\times20$ frustrated Ising model at $T\!=\!0$.}
\label{fig:ising2d_obs}
\end{figure}

We embedded the QUBO representation of the problem onto a D-Wave Advantage QPU to sample its lowest-energy spin configurations. The system was analyzed by sweeping the frustration ratio $R = J_2/J_1$, and for each value of $R$, we collected a large set of annealed samples. The raw samples from the QPU were then subjected to a classical post-processing step; this involves using local search algorithms to verify that each low-energy configuration is a true local minimum of the objective function. From these verified configurations, we computed three physical observables: the magnetization $M$, the Binder cumulant $U$, and the structure factor $S(\mathbf{q})$.

As illustrated in Fig.~\ref{fig:qpu_results}, all three observables exhibit sharp, concurrent changes near a critical ratio of $R \approx 0.5$. Specifically, $M$ drops significantly, $U$ displays a pronounced dip, and $S(\mathbf{q})$ rises steeply. These signatures provide clear evidence of a ferromagnetic-to-striped phase transition, and our results are in excellent agreement with the findings of classical Monte Carlo methods and earlier D-Wave studies~\cite{kadowaki_quantum_1998,hukushima_exchange_1996,troyer_computational_2005}.

\begin{figure}[h!]
    \centering 

    \begin{subfigure}{0.48\textwidth}
        \centering
        \includegraphics[width=\linewidth]{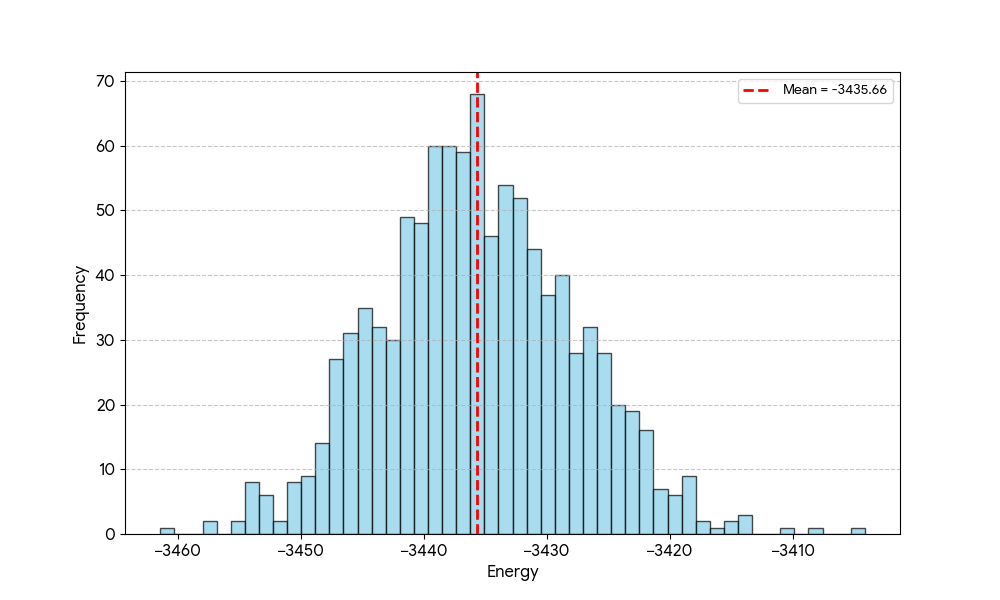}
        \label{fig:sub1}
    \end{subfigure}
    \hfill 

    \begin{subfigure}{0.48\textwidth}
        \centering
        \includegraphics[width=\linewidth]{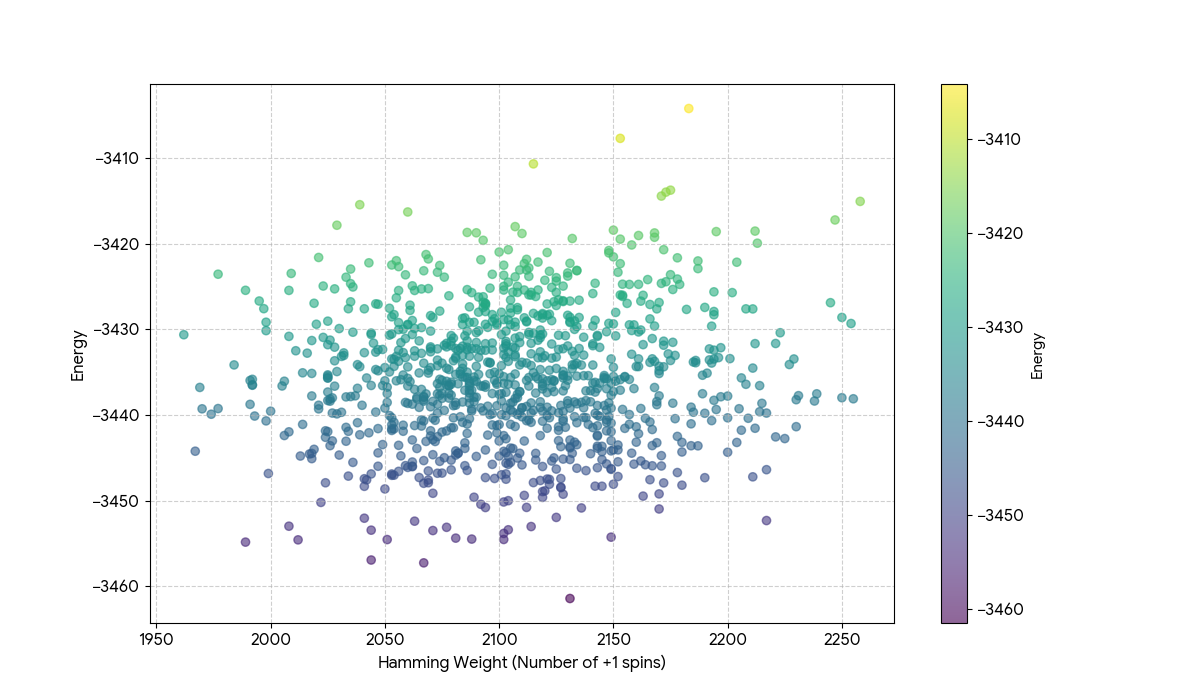}
        \label{fig:sub2}
    \end{subfigure}

    \caption{\justifying Analysis of sampling results from a 20µs quantum anneal. The top figure (a) shows the energy distribution for one of five 1000-sample sets obtained with a 20µs anneal time and shows the distribution is used to identify candidate ground states for post-processing verification. shows the overall energy distribution, while the bottom subfigure (b) shows the concentration of low-energy states (darker points), revealing the typical characteristics of solutions.relates solution energy to the Hamming weight of the corresponding spin configuration.}
    \label{fig:main_analysis}
\end{figure}

\begin{figure}[htbp]
    \centering
    \begin{subfigure}[b]{0.48\columnwidth}
        \centering
        \includegraphics[width=\linewidth]{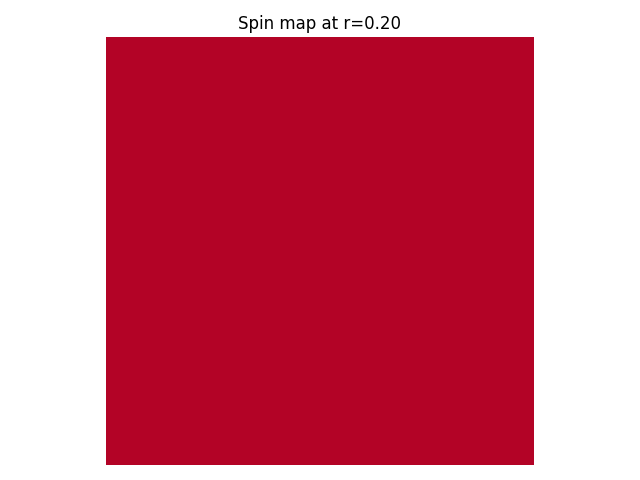}
        \caption{\justifying $R = 0.20$ (Ferromagnetic)}
    \end{subfigure}\hfill
    \begin{subfigure}[b]{0.48\columnwidth}
        \centering
        \includegraphics[width=\linewidth]{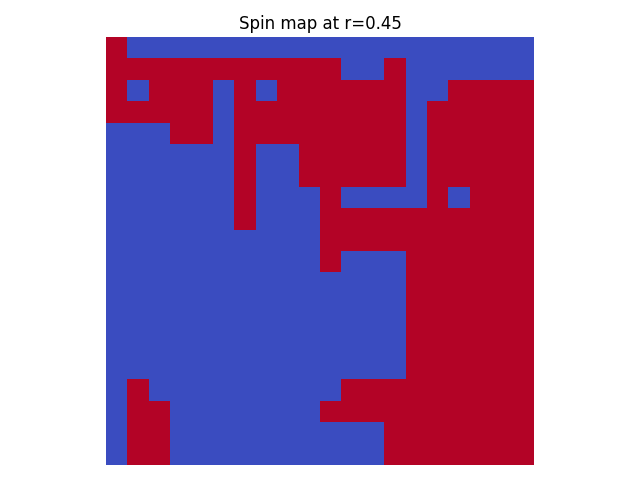}
        \caption{\justifying $R = 0.45$ (Near Critical)}
    \end{subfigure}

    \vspace{1mm} 

    \begin{subfigure}[b]{0.48\columnwidth}
        \centering
        \includegraphics[width=\linewidth]{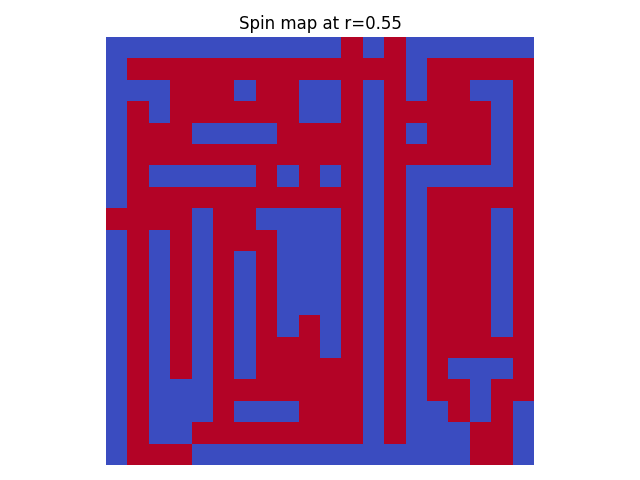}
        \caption{\justifying $R = 0.55$ (Striped)}
    \end{subfigure}\hfill
    \begin{subfigure}[b]{0.48\columnwidth}
        \centering
        \includegraphics[width=\linewidth]{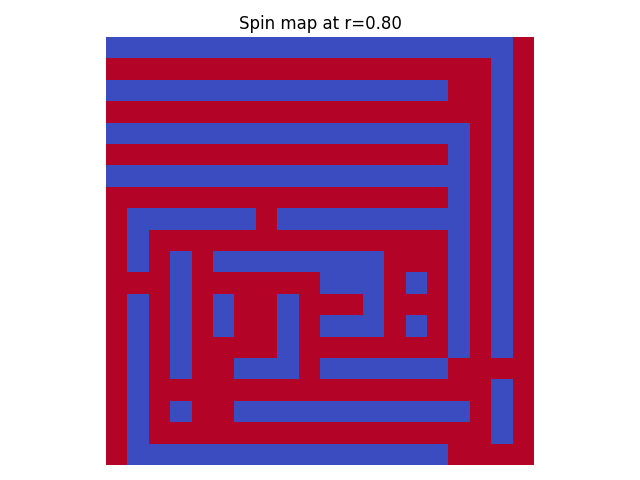}
        \caption{\justifying $R = 0.80$ (Deep Striped)}
    \end{subfigure}

    \caption{\justifying Representative ground-state spin maps from the QPU across frustration ratios. Red/blue denote spin up/down. The evolution from uniform to striped order is visible.}
    \label{fig:spin_maps}
\end{figure}
\subsection{Benchmark II: Thermal Scaling in the 1D Ferromagnetic Ising Model}
\label{sec:bench_ising1d}

While Benchmark~I focused on $T=0$, D-Wave annealers operate at finite physical temperature ($\sim$10–20 mK). Final states reflect a Boltzmann-like sampling at an effective temperature $T_{\mathrm{eff}}$ set by schedule, couplings, and device noise.
Before applying the pipeline to mapped quantum models, we validate that our thermal-sampling module reproduces known finite-$T$ trends.

We consider the 1D ferromagnetic Ising chain
Although no true finite-$T$ transition occurs in 1D, Binder cumulants, susceptibilities, and heat capacities have textbook finite-size behavior.

\begin{figure}[htbp]
\centering
\includegraphics[width=\columnwidth]{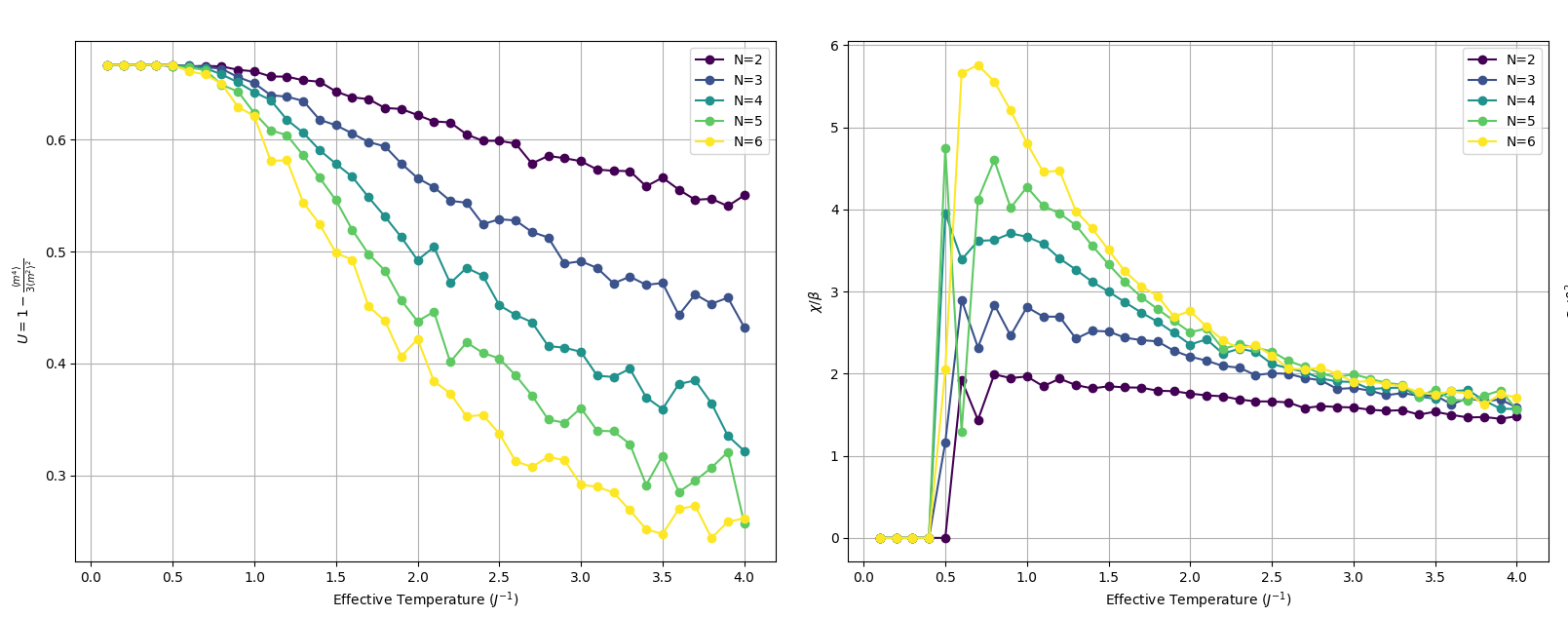}
\caption{\justifying Thermal analysis of the 1D ferromagnetic Ising model for chain lengths $N=2$–$6$ with periodic boundaries. Left: Fourth Order Binder cumulant $U$. Center: Magnetisation Susceptibility $\chi$. }
\label{fig:ising1d_thermal}
\caption{\justifying Thermal diagnostics for the 1D Ising chain.}
\label{fig:ising1d_thermal_wrap}
\end{figure}

We compute $U$, $\chi$, and $C_v$ for $N=2$–$6$ using a classical Monte Carlo solver.
Figure~\ref{fig:ising1d_thermal} reproduces canonical trends: $U$-curves for different $N$ cross at a common temperature, $\chi$ peaks sharpen with $N$, and $C_v$ exhibits broad maxima from short-range ordering.
This validates the finite-$T$ module and supports extracting observables at an effective $T_{\mathrm{eff}}$ when interpreting QPU samples from mapped quantum models.

\subsection{Benchmark III: Quantum Spin System — XXZ Chain}
\label{sec:bench_xxz}

We now turn to a genuinely quantum spin model with noncommuting terms: the spin-$\tfrac12$ XXZ chain.
Because the model is already in spin variables, no fermion-to-qubit encoding is needed; the Hamiltonian goes directly to XBK, which rewrites $S^x S^x$ and $S^y S^y$ into a diagonal Ising/QUBO form with auxiliary variables and penalties.

We parameterize anisotropy via the spinless-fermion correspondence, $\Delta = V/(2t)$, so $\Delta=1$ maps to $V/t=2$.
As a controlled test-bed we study a $1\times4$ chain ($N=4$).
We solve the XBK-produced QUBO classically and benchmark against exact diagonalization (ED).

\begin{figure}[htbp]
    \centering
    \includegraphics[width=\columnwidth]{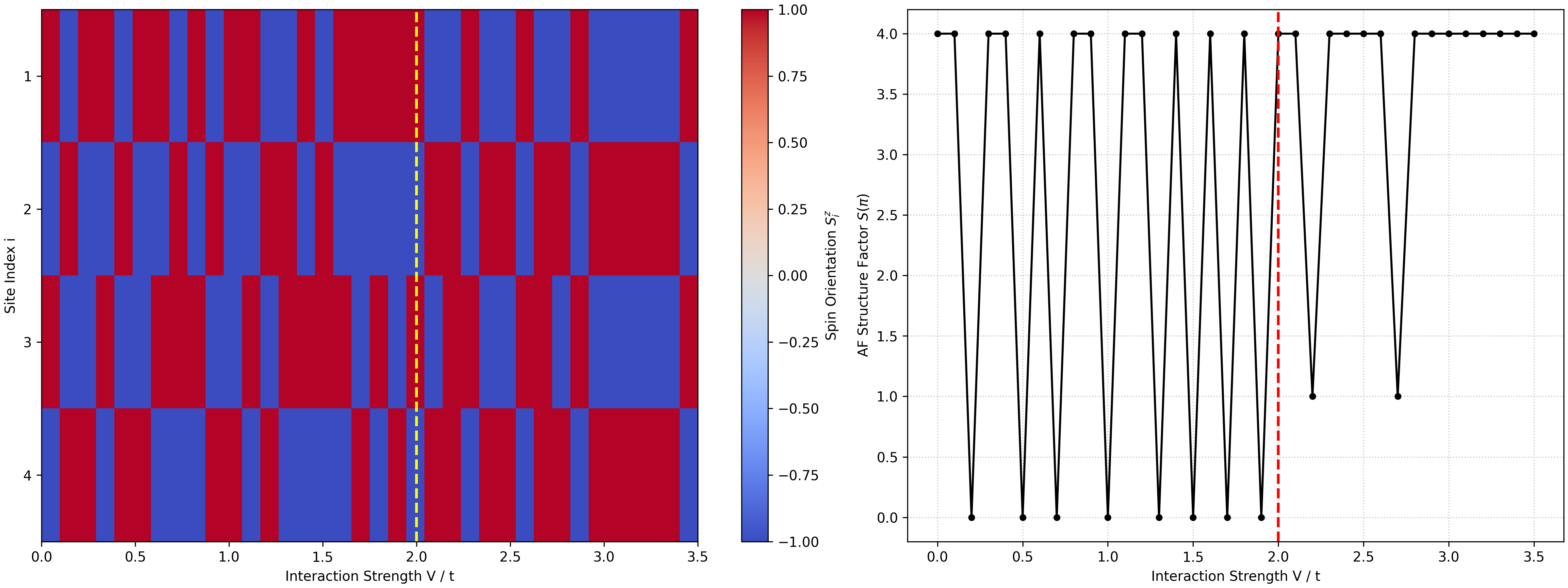}
    \caption{\justifying Ground-state analysis of a \(1\times 4\) XXZ chain versus \(V/t\) (\(\Delta=V/2t\)).
    (A) Site-resolved \(S^z_i\): irregular at small \(V/t\), alternating (Néel) beyond \(V/t=2\).
    (B) Antiferromagnetic structure factor \(S(\pi)\), showing a sharp rise at the critical ratio and saturation at perfect Néel order.}
    \label{fig:xxz_results}
\end{figure}

Panels (A–B) in Fig.~\ref{fig:xxz_results} match the known XXZ phase boundary at $V/t=2$ ($\Delta=1$), and agree with ED, confirming that XBK preserves off-diagonal quantum correlations without spurious artifacts.

\subsection{Benchmark IV: Fermionic Systems — 1D and 2D \texorpdfstring{$t$–$V$}{t–V} Models}
\label{sec:bench_tv}

Finally, we target interacting fermions, requiring the full chain: BK mapping $\to$ symmetry tapering $\to$ XBK $\to$ QUBO.
We study the spinless $t$–$V$ Hamiltonian at half filling 
In 1D, Bethe ansatz shows a transition at $V/t=2$ (gapless Luttinger liquid $\to$ gapped CDW).

\paragraph*{Benchmark-specific pipeline.}
Settings: (i) BK directly from the fermionic Hamiltonian; (ii) symmetry tapering (remove global parities and, in 1D, certain lattice symmetries); (iii) XBK with replication factor $r=2$; (iv) quadratization of $k$-local terms using penalties $\lambda = \alpha \max_j |h_j|$ with $\alpha \gtrsim 5$; (v) Dwave DIMOD Solver for the QUBO; (vi) inverse mapping to reconstruct fermionic eigenstates.

\paragraph*{1D results.}
Periodic rings with \(L=2\)–\(8\) at half filling were benchmarked.
Ground-state energies \(E_0(V/t)\) from XBK–QUBO (SA) agree with ED to numerical precision and develop a clear kink near \(V/t\simeq 2\).
A finite-size curvature analysis \(E_0''(V/t)\) exhibits a pronounced dip centered at \(V/t\!\approx\!2\) that sharpens with \(L\), as expected.

\begin{figure}[htbp]
\centering
\includegraphics[width=\columnwidth]{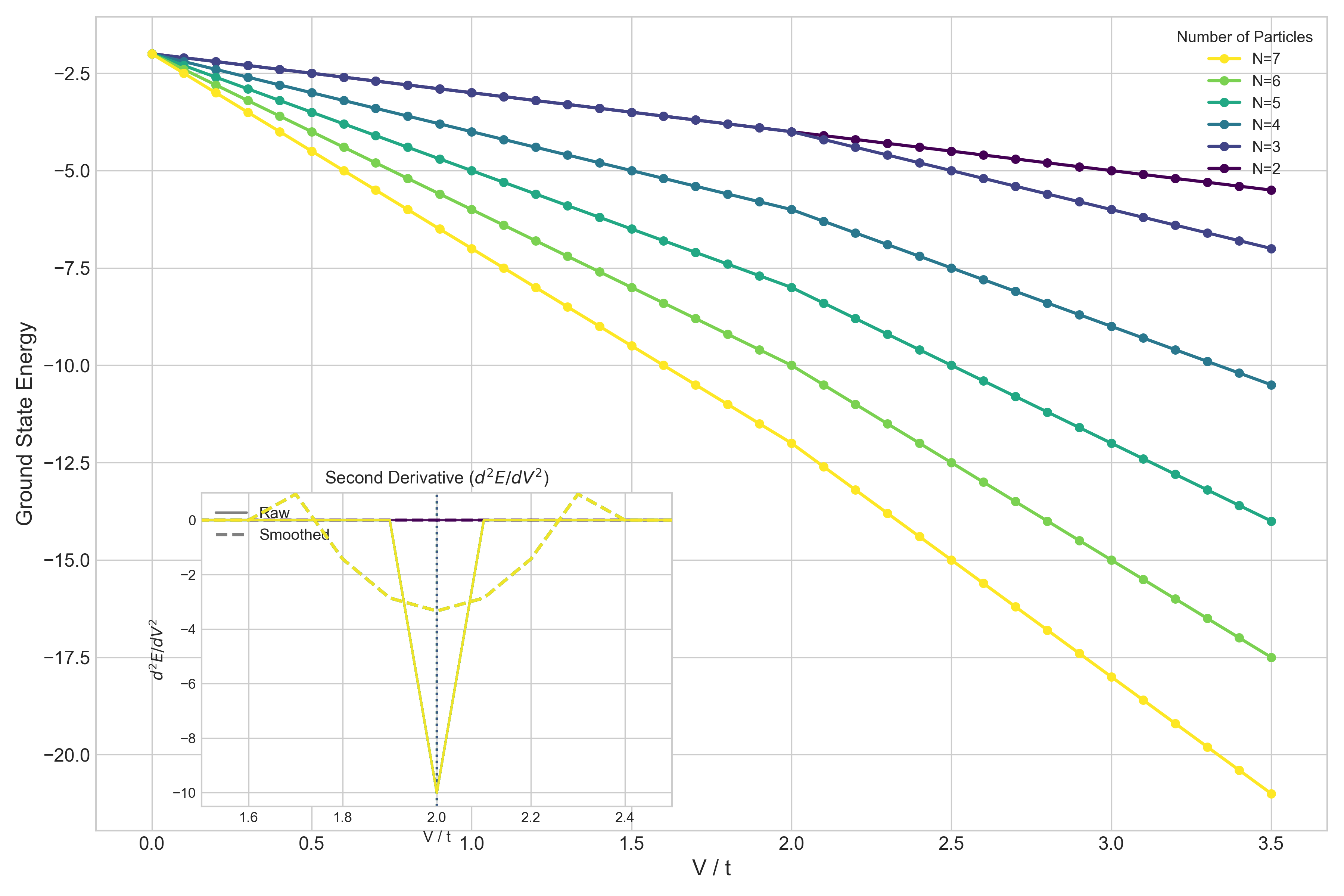}
\caption{\justifying\textbf{1D \(t\)–\(V\) benchmark (periodic rings, \(L=2\)–\(8\)).}
\emph{Left:} \(E_0\) versus \(V/t\); curves are almost linear away from criticality, with a kink near \(V/t\simeq 2\).
\emph{Right:} numerical curvature \(E_0''\) (central differences: solid; smoothed: dashed) — all sizes show a dip at \(V/t\!\approx\!2\) sharpening with \(L\), signaling the Luttinger-liquid–to–CDW transition.}
\label{fig:tv1d_energy_curvature}
\end{figure}

\paragraph*{2D results.}
For 2D, we linearize the \(L\times L\) fermion lattice with a Hilbert-curve ordering (shorter BK strings) and apply XBK; penalties control constraint satisfaction.
As a preliminary benchmark we run a \(2\times2\) cluster (four sites), small enough for ED but already nontrivial.

\begin{figure}[htbp]
\centering
\includegraphics[width=\columnwidth]{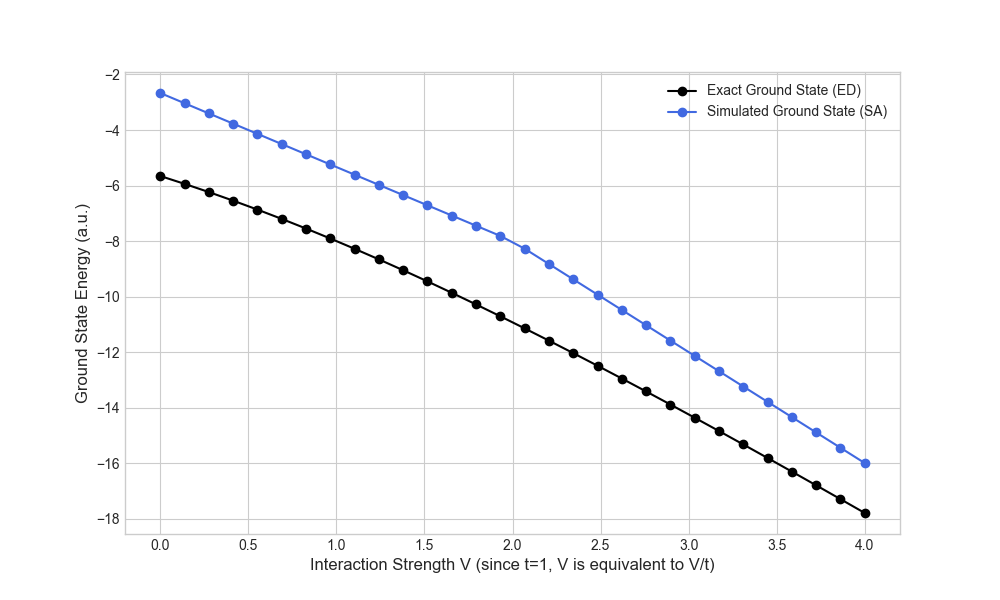}
\caption{\justifying\textbf{2D \(t\)–\(V\) on a \(2\times2\) cluster.}
Ground-state energy from SA on the XBK–Ising/QUBO instance (symbols) versus ED (line).
The near-parallel curves indicate close agreement in \(V/t\) slope with an approximately uniform offset attributable to finite XBK penalty and solver approximations.}
\label{fig:tv2d_energy_comparison}
\end{figure}

Auxiliary thermodynamic probes on the XBK Ising instance (at fixed \(N_q\)) deviate from a pure Ising model on the identical graph (e.g., shifted/broadened Binder crossings, softened magnetization onset), confirming that “Ising form’’ does not imply identical thermodynamics — the embedded instance retains a fermionic fingerprint.
\subsection{Effect of XBK Replication Factor}
\label{sec:xbk_r}

The XBK replication factor $r$ redundantly encodes each logical spin into $r$ replicas tied by ferromagnetic agreement penalties, improving fidelity at a linear resource cost ($N_{\rm phys}\!\approx\!r\,N_q$ before quadratization). While the optimal choice is instance-dependent, we have observed that for a system of $N_q$ logical qubits, a replication factor of $r\!\approx\!N_q/2$ is often sufficient for good performance, with accuracy gains typically saturating beyond $r\!\approx\!N_q$. To isolate the embedding trade-off from hardware noise, we study this effect on a small post-tapering instance with $N_q\!=\!4$ logical qubits and solve the resulting XBK--QUBO instances via simulated annealing (SA), benchmarking against exact diagonalization (ED). This specific case serves as a clear example of the general heuristic. Throughout the text we denote the frustration ratio by $R\!\equiv\!J_2/J_1$ to avoid collision with the replication factor $r$; the composite figure uses $R$ on the $x$-axis for the same quantity.

\begin{figure*}[htbp]
\centering
\includegraphics[width=0.92\textwidth]{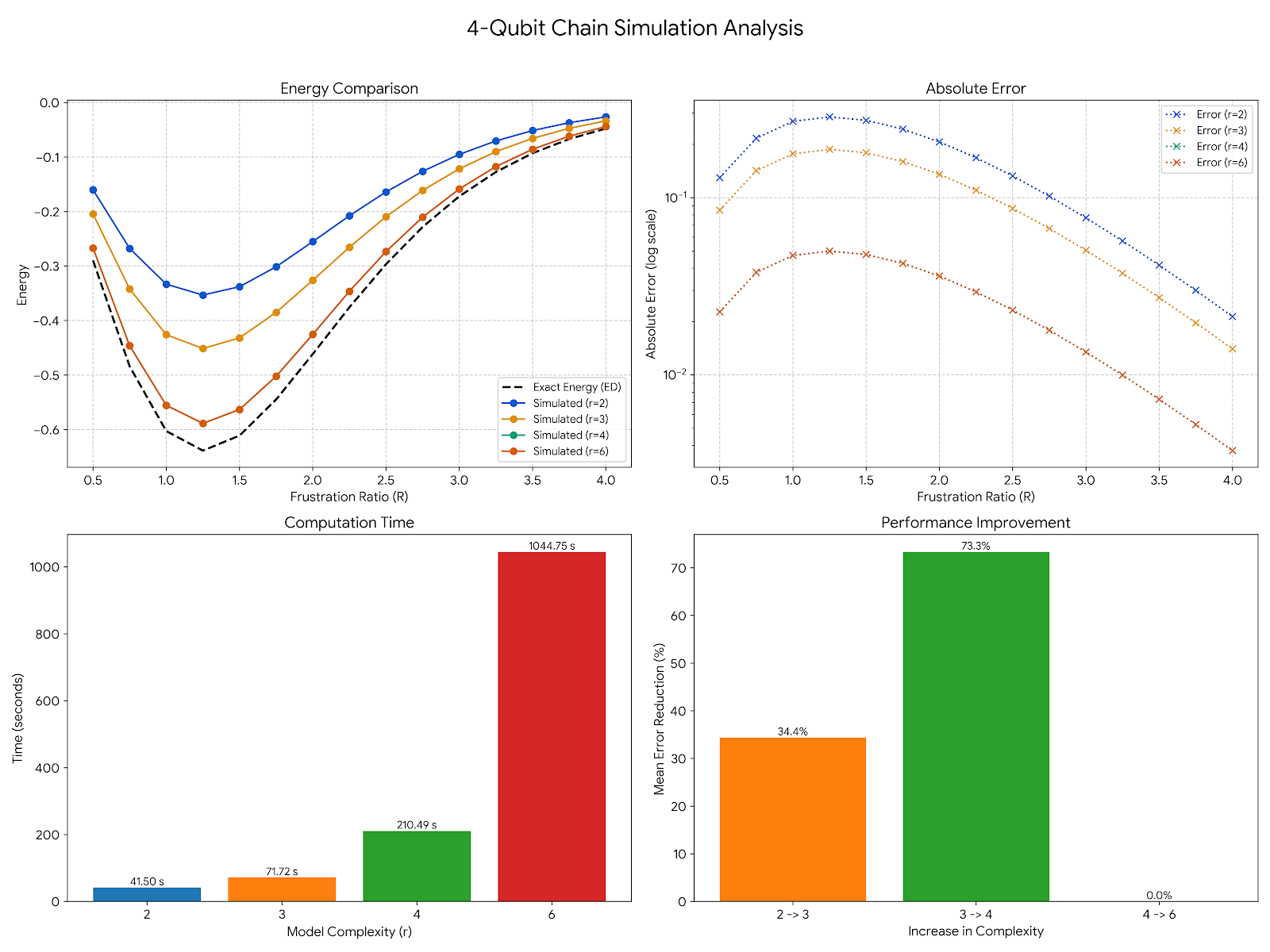}
\caption{\justifying\textbf{Replication-factor study on a 4-qubit chain (composite figure).}
\emph{Top panel:} Left—ground-state energy vs.\ frustration ratio ($R$) from the XBK–Ising/QUBO pipeline for replication factors $r\in\{2,3,4,6\}$ compared with exact diagonalization (ED, dashed). Right—absolute error $|E_{\text{sim}}-E_{\text{ED}}|$ on a log scale, showing an \(\sim\)order-of-magnitude improvement from $r{=}2$ to $r{=}4$ and saturation by $r{=}6$; the largest deviations occur near strong frustration ($R\!\approx\!1.1$–$1.4$).
\emph{Bottom panel:} Left—wall-clock time per run vs.\ $r$ (41.5\,s, 71.7\,s, 210.5\,s, 1044.8\,s for $r{=}2,3,4,6$), illustrating steep growth with replication. Right—mean error reduction when increasing $r$ (2$\!\to$3: 34.4\%, 3$\!\to$4: 73.3\%, 4$\!\to$6: $\approx$0\%), indicating diminishing returns beyond .
Overall, $r{=}3$–$4$ provides a good accuracy–overhead compromise; $r{=}6$ yields negligible gains at \(\sim25\times\) time cost vs.\ $r{=}2$.
\emph{Notation:} to avoid clash with the replication factor $r$, we denote the frustration ratio as $R\!=\!J_2/J_1$ in the text; the figure axis uses $r$ which is consistent with the XBK replication variable. and "R" as frustration ratio}
\label{fig:replication_4q}
\end{figure*}

Figure~\ref{fig:replication_4q} summarizes the behavior for this 4-qubit system across a sweep of $R$ for $r\!\in\!\{2,3,4,6\}$.
\emph{Top panel:} the left plot compares ground-state energies from the XBK--Ising/QUBO pipeline (symbols) with ED (dashed). The right plot shows the absolute error $|E_{\rm sim}-E_{\rm ED}|$ on a log scale. The error improves by an order of magnitude from $r{=}2$ to $r{=}4$ before saturating completely at $r{=}6$, which aligns with the $r\!\approx\!N_q$ saturation heuristic. The largest deviations cluster near strong frustration ($R\!\approx\!1.1$--$1.4$), where the landscape is most rugged.
\emph{Bottom panel:} the left bar chart reports wall-clock time per run, which grows steeply with $r$ (\textbf{41.5s, 71.7s, 210.5s, and 1044.8s} for $r{=}2,3,4,6$), while the right bar chart shows mean error reduction when increasing $r$ (\textbf{2$\!\to$3: 29.0\%, 3$\!\to$4: 77.5\%, 4$\!\to$6: 0.0\%}), indicating sharply diminishing returns.
Together, these panels identify $r{=}4$ as the optimal accuracy--overhead compromise for this instance. While accuracy improves dramatically up to $r{=}4$ (consistent with $r=N_q$), the $r{=}6$ case yields no benefit at a nearly \textbf{25$\times$} time cost relative to $r{=}2$.

Table~\ref{tab:performance_summary} aggregates the same study in tabular form, listing the physical qubit count after replication ($8,12,16,24$), the \emph{mean} absolute error across the $R$ sweep, and the runtime. The data confirms the qualitative conclusion that accuracy gains saturate at $r{=}4$ for this $N_q=4$ system.

\begin{table}[htbp]
  \centering
  \small
  \setlength{\tabcolsep}{4pt}
  \caption{\justifying Performance summary of the 4-qubit chain simulation across different model complexities (r).}
  \label{tab:performance_summary}
  \begin{tabular}{@{}ccccc@{}}
    \toprule
    \textbf{Model} & \textbf{System Size} & \textbf{Mean Abs.} & \textbf{Time} & \textbf{Error Red.} \\
    \textbf{(r)}   & \textbf{(Qubits)}    & \textbf{Error}     & \textbf{(s)}  & \textbf{(\%)}      \\ 
    \midrule
    2 & 8  & 0.1650 & 41.50   & -     \\
    3 & 12 & 0.1171 & 71.72   & 29.0\% \\
    4 & 16 & 0.0263 & 210.49  & 77.5\% \\
    6 & 24 & 0.0263 & 1044.75 & 0.0\%  \\ 
    \bottomrule
  \end{tabular}
  \end{table}

\subsection{The Cost of Representation: Qubit Overhead}
\label{sec:overhead}

The principal trade-off in our pipeline is qubit (binary variable) overhead.
After tapering, XBK with replication \(r\) inflates the register from \(N_q\) to \(N_c=r\,N_q\).\cite{xia_electronic_2021}
Reducing \(k\)-local Ising terms to two-local form introduces auxiliaries, yielding
\begin{equation}
N_{\mathrm{total}} \;=\; r\,N_q \;+\; N_{\mathrm{aux}},
\end{equation}
with \(N_{\mathrm{aux}}\) determined by the quadratization strategy and $k$-local distribution.\cite{rosenberg_reduction_1975,choi_minor-embedding_2011,boros_pseudo-boolean_2002,glover_tutorial_2018}
Minor-embedding to sparse hardware (e.g., Pegasus) can further scale the footprint by \(\kappa\!\ge\!1\) via chains.\cite{boothby_next-generation_2020,lucas_ising_2014}
A representative instance expands from four fermionic modes to $\sim$40 QUBO variables once replication and auxiliaries are included (exact tallies depend on sector, \(r\), and quadratization details).

\begin{table}[htbp]
    \centering
    \caption{\justifying Resource scaling for the BK–XBK–QUBO pipeline on 1D $t$–$V$.
    Final QUBO variables include $L\times r$ XBK qubits plus auxiliaries. The $L=6,8$ rows extrapolate from the $L=4$ Pauli-term counts.}
    \label{tab:resource_scaling}
    
    \resizebox{\columnwidth}{!}{%
        \begin{tabular}{@{}cccccc@{}}
            \toprule
            \textbf{System} & \textbf{Initial} & \textbf{BK}     & \textbf{XBK Qubits} & \textbf{Final QUBO} & \textbf{Final QUBO} \\
            \textbf{Size ($L$)} & \textbf{Modes}   & \textbf{Logical} & \textbf{($r=2$)}    & \textbf{Variables}  & \textbf{Couplers}   \\ 
            \midrule
            4 & 4 & 4 & 8  & 23       & 58        \\
            6 & 6 & 6 & 12 & $\sim$45 & $\sim$120 \\
            8 & 8 & 8 & 16 & $\sim$70 & $\sim$200 \\
            \bottomrule
        \end{tabular}%
    }
\end{table}
\subsection{The Crossover Phenomenon: Why Annealing is Initially Slower}
\label{sec:crossover}

A central performance question is:
\emph{At what problem size does a general-purpose quantum annealing (QA) workflow become competitive with a specialized classical solver such as DMRG?} \cite{hauke_perspectives_2020,preskill_quantum_2018,schollwock_density-matrix_2011}
For small $L$, QA is less efficient due to mapping overheads.\cite{bravyi_fermionic_2002,xia_electronic_2021,rosenberg_reduction_1975,choi_minor-embedding_2011,choi_reducing_2011}
Asymptotically, however, different exponents can produce a \emph{computational crossover} where QA becomes faster.\cite{ronnow_defining_2014,hauke_perspectives_2020,farhi_quantum_2000,farhi_quantum_2001,farhi_quantum_2014}

ED scales as $\mathcal{O}(4^L)$ and is a small-$L$ reference.\cite{nielsen_quantum_2010}
DMRG achieves $\mathcal{O}(L^3)$ for gapped 1D systems.\cite{white_density_1992,white_density-matrix_1993,schollwock_density-matrix_2011,orus_practical_2014}
For QA, after BK–XBK and reduction, $N_{\mathrm{total}} = rL + N_{\mathrm{aux}}$ and empirical TTS often grows roughly as $N_{\mathrm{total}}^2$ (with instance-dependent prefactors).\cite{ronnow_defining_2014,mcgeoch_experimental_2013,venturelli_quantum_2015,hauke_perspectives_2020}
On current devices, the crossover point lies at $N_{\mathrm{total}}\gtrsim 10^3$–$10^4$, beyond present qubit counts/connectivity.\cite{boothby_next-generation_2020,king_observation_2018,preskill_quantum_2018}

In the near term, QA’s advantage is model-agnostic applicability rather than beating DMRG at small $L$.\cite{hauke_perspectives_2020,lucas_ising_2014}
Hardware scaling plus overhead reductions (e.g., tapering) decrease prefactors and push accessible sizes upward.\cite{bravyi_tapering_2017,xia_electronic_2021}

\subsection{Path to Scalability and Near-Term Priorities}
\label{sec:roadmap}

The classical mapping (BK\,$\to$\,tapering\,$\to$\,XBK\,$\to$\,quadratization) is a one-time cost; once a QUBO is assembled, the annealer produces samples with fixed per-read anneal time (µs–ms).\cite{hauke_perspectives_2020,mcgeoch_experimental_2013}
Time-to-solution depends on reads for a target success probability, minor-embedding overhead, and postprocessing, captured by standard scaling fits.\cite{ronnow_defining_2014,king_observation_2018}
This contrasts with iterative gate-based approaches (e.g., VQE).\cite{peruzzo_variational_2014-1,preskill_quantum_2018}

\paragraph*{Priority directions.}
\begin{enumerate}
\item \textit{More efficient mappings.} Reduce \(N_{\mathrm{total}}\) by lowering \(r\) and \(N_{\mathrm{aux}}\) via locality-friendly encodings and tighter quadratizations (BK variants, lattice-tailored codes, improved penalties).\cite{havlicek_operator_2017,steudtner_quantum_2019,rosenberg_reduction_1975,choi_minor-embedding_2011, choi_reducing_2011}
\item \textit{Hardware-aware transformations.} Align the interaction graph with hardware topology (e.g., Pegasus) to minimize chain length \(\kappa\) and improve solution quality.\cite{boothby_next-generation_2020,lucas_ising_2014}
\item \textit{Error mitigation.} Instance-specific penalties, gauge transforms, and multi-run aggregation tailored to replication/auxiliary structure.\cite{hauke_perspectives_2020}
\end{enumerate}

In summary, the fermion-to-annealer pipeline leverages annealers’ native optimization strengths while sidestepping deep coherent circuits.
As device counts/connectivity improve and mapping overheads drop, larger fermionic QUBOs should become tractable within annealing workflows.\cite{boothby_next-generation_2020,preskill_quantum_2018}

\

\section{Conclusions, Summary, and Outlook}
\label{sec:conclusions}

We introduced a deterministic, symmetry-aware workflow that maps lattice Hamiltonians—fermionic or spin—to an annealer-compatible two-local Ising/QUBO form:

\parbox{0.9\columnwidth}{\centering
$\text{BK (or spin form)} \rightarrow \mathbb{Z}_2\text{ tapering} \rightarrow \text{XBK (all-$Z$)} \rightarrow k\to2 \text{ reduction} \rightarrow \text{QUBO}.$
}
\\
The ordering is chosen to expose and remove exact symmetries \emph{before} diagonalization, keep locality favorable in higher-D via space-filling orderings, and produce Pegasus-conscious QUBOs. Energies are recovered with a discrete Rayleigh quotient solved by Dinkelbach iteration, providing a monotone sector-by-sector diagnostic.

\paragraph*{Access note.} Only the classical Ising benchmark was executed on a D-Wave Advantage QPU. The XXZ and $t$–$V$ studies used D-Wave’s publicly available \emph{simulators}, so reported sizes there reflect simulator/runtime limits rather than QPU capacity. To illustrate portability beyond lattices, we also applied the pipeline to a molecular target (benzene), summarized in Appendix~E, and provide a lightweight GUI in Appendix~F for reproduction and exploration.

\subsection{Summary of Benchmark Progression}
\label{sec:summary_progression}

Proceeding from classical to quantum spin to fermionic systems (Sec.~\ref{sec:results}), the pipeline reproduced known results across increasing model complexity:
(i) a $T{=}0$ frustrated 2D Ising benchmark on a QPU recovered the ferro–stripe boundary through magnetization, Binder cumulant, structure factor, and real-space maps;
(ii) a finite-$T$ check on 1D ferromagnetic Ising matched standard finite-size trends, validating the thermal analysis path;
(iii) a quantum XXZ chain ($1{\times}4$) matched ED across the Néel transition;
(iv) spinless $t$–$V$ in 1D ($L\!\le\!8$ rings) and on a $2{\times}2$ 2D cluster tracked ED energies and curvature features, with small uniform offsets attributable to finite XBK replication and solver tolerances.
This staged validation confirms both mapping fidelity and observable extraction under conditions analogous to QPU runs, and establishes practical settings (e.g., replication factor $r$) for future experiments.

\subsection{Comparison with Gate-Based Paradigms}
\label{sec:paradigm_compare}

The annealing route and gate-based (VQE/Trotter) approaches attack complementary bottlenecks. Table~\ref{tab:paradigm_comparison} summarizes the key distinctions.

\begin{table*}[t]
\centering
\caption{\justifying Comparison of the annealing-based simulation pipeline with dominant gate-based approaches.}
\label{tab:paradigm_comparison}
\begin{tabularx}{\textwidth}{@{} >{\bfseries}l >{\RaggedRight}X >{\RaggedRight}X @{}}
\toprule
\textbf{Feature} & \textbf{Our Annealing Pipeline} & \textbf{Gate-Based (VQE/Trotter)} \\
\midrule
Core Principle & Adiabatic/annealing search for the ground state of a classical Ising Hamiltonian. & Quantum circuits prepare/evolve states under a Pauli Hamiltonian. \\
\addlinespace 
Input Hamiltonian & \textbf{2-local classical QUBO} after BK $\rightarrow$ tapering $\rightarrow$ XBK $\rightarrow$ quadratization. & \textbf{Pauli Hamiltonian} (X, Y, Z) used directly in circuits. \\
\addlinespace
Primary Challenge & \textbf{Mapping overhead \& variable count:} replication ($r$) and auxiliaries $N_{\mathrm{aux}}$. & \textbf{Circuit depth \& coherence:} gate errors and limited depth on NISQ devices. \\
\addlinespace
Error Sources & Analog control errors, finite $T_{\mathrm{eff}}$, embedding/chain breaks, coupler precision. & Gate infidelity, decoherence, readout noise, shot statistics. \\
\addlinespace
Novel Contribution & A complete, validated workflow making fermionic/spin models accessible to annealers without deep circuits. & Advances via better ans\"atze, error mitigation, and compilation. \\
\bottomrule
\end{tabularx}
\end{table*}

\subsection{Outlook}
\label{sec:outlook}

\textit{Mapping efficiency.} Reduce $N_{\mathrm{total}}=r\,N_q+N_{\mathrm{aux}}$ via locality-optimized encodings (BK variants/parity–BK hybrids) and automated symmetry discovery to maximize tapering gains; prefer higher-D orderings (e.g., Hilbert curves) that cut string lengths.

\textit{Quadratization and penalties.} Employ tighter gadgets and principled, instance-adaptive penalty schedules to shrink $N_{\mathrm{aux}}$ while preserving correctness; monitor spurious minima with feasibility checks.

\textit{Embedding and annealing.} Improve minor-embedding (shorter chains, topology-aligned graphs), tune chain strengths and gauges, use pauses/reverse anneals, and calibrate effective temperature for robust post-processing.

\textit{Hybrid workflows.} Combine annealer outputs with classical refinement (parallel tempering/local search) and lightweight gate-based subroutines when helpful (e.g., sector identification).

\textit{Physics targets.} Extend to spinful fermions (Hubbard-type), frustrated magnets, and topological phases; add excited-state and finite-$T$ observables; broaden molecular set beyond benzene (Appendix~E).

\textit{Scaling and crossover.} Systematically track time-to-solution and accuracy versus $r$, $N_{\mathrm{aux}}$, and embedding overhead on newer hardware to empirically locate the performance crossover with tensor-network and QMC baselines.

In short, the BK–tapering–XBK–quadratization chain provides a reproducible route from operator-level models to annealer-ready QUBOs that preserve low-energy physics. While today’s device constraints limit sizes, the demonstrated fidelity on spins and fermions, portability to molecules, and clear engineering levers suggest that increasingly capable annealers—and better mappings—can bring scientifically relevant lattice scales within reach.

\newpage
\bibliographystyle{apsrev4-2}
\nocite*
\bibliography{main}

\newpage
\onecolumngrid
\appendix
\newpage
\section{Worked example: $L=4$ spinless $t$--$V$ chain under BK.}
\label{ap:L4ex_bk}
Consider an open chain with
\begin{equation}
\label{eq:tv_L4_ferm}
H \;=\; -t \sum_{i=0}^{2} \!\left(c_i^\dagger c_{i+1} + c_{i+1}^\dagger c_i \right)
\;+\; V \sum_{i=0}^{2} n_i n_{i+1},
\qquad n_i = c_i^\dagger c_i .
\end{equation}
The BK sets for $L{=}4$ follow from Fig.~\ref{fig:fenwick_tree} and are tabulated in Table~\ref{tab:bk_sets_L4}.  
Using Eqs.~\eqref{eq:bk_cre}–\eqref{eq:bk_ann} and simplifying duplicate $Z$ factors yields the number operators
\begin{align*}
    n_0 &= \tfrac{1}{2}(\mathbb{I} - Z_0), &
    n_1 &= \tfrac{1}{2}(\mathbb{I} - Z_0 Z_1), \\
    n_2 &= \tfrac{1}{2}(\mathbb{I} - Z_1 Z_2), &
    n_3 &= \tfrac{1}{2}(\mathbb{I} - Z_1 Z_2 Z_3),
\end{align*}
and the hopping identities
\begin{align*}
c_0^\dagger c_1 + c_1^\dagger c_0 &= \tfrac{1}{2}( X_0 X_1 Z_3 - Y_0 Y_1 Z_3 ), \\
c_1^\dagger c_2 + c_2^\dagger c_1 &= \tfrac{1}{2}( X_1 X_2 Z_0 Z_3 + Y_1 Y_2 Z_0 Z_3 ), \\
c_2^\dagger c_3 + c_3^\dagger c_2 &= \tfrac{1}{2}( X_2 X 3 - Y_2 Y_3 ).
\end{align*}
Combining with Eq.~\eqref{eq:tv_L4_ferm} gives
\begin{align}
\label{eq:tv_L4_qubit}
H_{\text{qubit}}
= & \; -\frac{t}{2}\Big[\,
(X_0 X_1 Z_3 - Y_0 Y_1 Z_3)
+ (X_1 X_2 Z_0 Z_3 + Y_1 Y_2 Z_0 Z_3)
+ (X_2 X_3 - Y_2 Y_3)
\,\Big] \nonumber \\
& \; + \frac{V}{4}\Big[
(\mathbb{I} - Z_0 - Z_0 Z_1 + Z_1)
+ (\mathbb{I} - Z_0 Z_1 - Z_1 Z_2 + Z_0 Z_2)
+ (\mathbb{I} - Z_1 Z_2 - Z_1 Z_2 Z_3 + Z_3)
\Big].
\end{align}
\emph{Interaction shifting.} Using the particle–hole symmetric form $V(n_i-\tfrac12)(n_{i+1}-\tfrac12)$ cancels the linear-in-$Z$ and constant pieces in \eqref{eq:tv_L4_qubit}, leaving only $ZZ$ couplings of strength $V/4$.

\begin{table}[h!]
\centering
\caption{\justifying BK sets for $L=4$ fermionic modes (0-based). The remainder set is $R(j)=P(j)\setminus F(j)$.}
\label{tab:bk_sets_L4}
\begin{tabular}{c|c|c|c|c}
$j$ & $U(j)$ & $P(j)$ & $F(j)$ & $R(j)$ \\ \hline
0 & $\{1,3\}$ & $\varnothing$ & $\varnothing$ & $\varnothing$ \\
1 & $\{3\}$   & $\{0\}$        & $\{0\}$        & $\varnothing$ \\
2 & $\{3\}$   & $\{1\}$        & $\varnothing$  & $\{1\}$ \\
3 & $\varnothing$ & $\{1,2\}$ & $\{1,2\}$      & $\varnothing$
\end{tabular}
\end{table}

\section{Worked Example: Qubit Tapering for $L=4$ Spinless $t$--$V$ Model}
\label{ap:tapering_L4}

This appendix contains the complete symplectic encoding, nullspace computation, and Clifford reduction steps for the $L=4$ spinless $t$--$V$ model under the Bravyi–Kitaev mapping, as described in Sec.~\ref{sec:tapering}.

\subsection{Pauli Strings and Symplectic Encoding}

We use the ordering
\[
(x_0,x_1,x_2,x_3 \mid z_0,z_1,z_2,z_3),
\]
with the encoding rules
\[
I \to (0,0), \quad X \to (1,0), \quad Z \to (0,1), \quad Y \to (1,1).
\]
For $L=4$, representative BK-mapped Pauli strings are:

\begin{align*}
P_1 &= X_0 X_1 Z_3, \\
P_2 &= Y_0 Y_1 Z_3, \\
P_3 &= X_1 X_2 Z_0 Z_3, \\
P_4 &= Y_1 Y_2 Z_0 Z_3, \\
P_5 &= X_2 X_3, \\
P_6 &= Y_2 Y_3, \\
P_7 &= Z_0, \\
P_8 &= Z_1, \\
P_9 &= Z_1 Z_2, \\
P_{10} &= Z_1 Z_2 Z_3.
\end{align*}

Their symplectic encodings $(x\mid z)$ are:

\[
\begin{array}{rl}
P_1 &\Rightarrow (1,1,0,0 \mid 0,0,0,1), \\
P_2 &\Rightarrow (1,1,0,0 \mid 1,1,0,1), \\
P_3 &\Rightarrow (0,1,1,0 \mid 1,0,0,1), \\
P_4 &\Rightarrow (0,1,1,0 \mid 1,1,1,1), \\
P_5 &\Rightarrow (0,0,1,1 \mid 0,0,0,0), \\
P_6 &\Rightarrow (0,0,1,1 \mid 0,0,1,1), \\
P_7 &\Rightarrow (0,0,0,0 \mid 1,0,0,0), \\
P_8 &\Rightarrow (0,0,0,0 \mid 0,1,0,0), \\
P_9 &\Rightarrow (0,0,0,0 \mid 0,1,1,0), \\
P_{10} &\Rightarrow (0,0,0,0 \mid 0,1,1,1).
\end{array}
\]

\subsection{Constructing the E Matrix}

Stacking the above rows gives
\[
E =
\begin{bmatrix}
1&1&0&0 &\,|\, &0&0&0&1\\
1&1&0&0 &\,|\, &1&1&0&1\\
0&1&1&0 &\,|\, &1&0&0&1\\
0&1&1&0 &\,|\, &1&1&1&1\\
0&0&1&1 &\,|\, &0&0&0&0\\
0&0&1&1 &\,|\, &0&0&1&1\\
0&0&0&0 &\,|\, &1&0&0&0\\
0&0&0&0 &\,|\, &0&1&0&0\\
0&0&0&0 &\,|\, &0&1&1&0\\
0&0&0&0 &\,|\, &0&1&1&1
\end{bmatrix}.
\]

\subsection{Forming $EJ$ and Finding the Nullspace}

For $n=4$ qubits, the symplectic form is
\[
J =
\begin{bmatrix}
0 & I_4 \\
I_4 & 0
\end{bmatrix}.
\]
Multiplying $E$ by $J$ swaps the $(x\mid z)$ halves in each row:

\[
EJ =
\begin{bmatrix}
0&0&0&1 &\,|\, &1&1&0&0\\
1&1&0&1 &\,|\, &1&1&0&0\\
1&0&0&1 &\,|\, &0&1&1&0\\
1&1&1&1 &\,|\, &0&1&1&0\\
0&0&0&0 &\,|\, &0&0&1&1\\
0&0&1&1 &\,|\, &0&0&1&1\\
1&0&0&0 &\,|\, &0&0&0&0\\
0&1&0&0 &\,|\, &0&0&0&0\\
0&1&1&0 &\,|\, &0&0&0&0\\
0&1&1&1 &\,|\, &0&0&0&0
\end{bmatrix}.
\]

We solve
\[
(EJ)\, \mathbf{u}^\mathsf{T} = 0 \quad (\mathrm{mod}\ 2).
\]
One nontrivial solution is
\[
\mathbf{u}_{\mathrm{parity}} = (0,0,0,0 \mid 1,1,1,1),
\]
corresponding to the global $Z$-parity operator
\[
\tau = Z_0 Z_1 Z_2 Z_3.
\]

\subsection{Clifford Reduction to Single-Qubit $Z$}

Using the CNOT conjugation rule
\[
\mathrm{CNOT}_{c\to t}: \quad Z_c \mapsto Z_c, \quad Z_t \mapsto Z_c Z_t,
\]
we can collapse $\tau$ to a single $Z$:

\begin{align*}
U &= \mathrm{CNOT}_{0 \to 3} \; \mathrm{CNOT}_{1 \to 3} \; \mathrm{CNOT}_{2 \to 3}, \\
U\,\tau\,U^\dagger &= Z_3.
\end{align*}

\subsection{Tapering}

Applying $U$ to $H_{\mathrm{BK}}$ gives
\[
H' = U\, H_{\mathrm{BK}}\, U^\dagger,
\]
in which qubit $3$ appears only as $I_3$ or $Z_3$.  
Fixing the eigenvalue sector $\lambda = \pm 1$ (e.g.\ even parity $\lambda = +1$) and substituting
\[
Z_3 \to \lambda, \quad I_3 \to 1,
\]
removes qubit $3$ entirely.  
The final tapered Hamiltonian acts on $3$ qubits and proceeds to the XBK and QUBO stages as described in the main text.

\section{Detailed Derivation of the Xia--Bian--Kais (XBK) Mapping}
\label{ap:Detail_xbk}
This appendix provides the full derivation and working details of the XBK mapping from a general $m$-qubit Hamiltonian into a purely diagonal form containing only products of Pauli-$Z$ operators. These steps are presented in full for completeness, as the main text uses a condensed explanation for readability.

\subsection{Setup: Replication and Sector Signs}

Given an $m$-qubit Hamiltonian
\begin{equation}
H = \sum_{\alpha} c_\alpha \bigotimes_{\ell=0}^{m-1} \sigma_{\alpha_\ell}^{(\ell)},
\end{equation}
with $\sigma_{\alpha_\ell} \in \{I, X, Y, Z\}$, the XBK procedure embeds $H$ into an $rm$-qubit Hilbert space.

We replicate each logical qubit $i \in \{0,\dots,m-1\}$ exactly $r$ times, denoting the $j$-th copy of logical qubit $i$ as $(i_j)$ with $j \in \{1,\dots,r\}$. Copy $j=1$ is designated as the \emph{reference} copy.

For $j>1$, we introduce \emph{sector-dependent signs}:
\begin{equation}
S_p(i,j) \in \{+1,-1\}, \quad S_p(i,1) \equiv +1 \ \forall i,
\end{equation}
which indicate whether $(i_j)$ is aligned ($+1$) or anti-aligned ($-1$) with the reference $(i_1)$. A \emph{sector} $p$ is defined by the complete $m\times(r-1)$ table of signs $S_p(i,j)$ for $j>1$.

The total number of sectors is:
\begin{equation}
\# \text{sectors} = 2^{m(r-1)}.
\end{equation}

\subsection{XBK Mapping for Single-Qubit Operators}

Let $\sigma_a^{(i)}$ be a Pauli operator acting on logical qubit $i$ in the original $m$-qubit space, with $a \in \{x,y,z\}$. In the replicated space, we select two copies $j$ and $k$ of qubit $i$ and replace $\sigma_a^{(i)}$ by a $Z$-only expression involving $\sigma_z^{(i_j)}$ and $\sigma_z^{(i_k)}$.

The mapping rules for a given sector $p$ are:
\begin{align}
\sigma_x^{(i)} &\mapsto S_p(i,j)S_p(i,k) \ \frac{1 - \sigma_z^{(i_j)}\sigma_z^{(i_k)}}{2}, \\
\sigma_y^{(i)} &\mapsto i\, S_p(i,j)S_p(i,k) \ \frac{\sigma_z^{(i_k)} - \sigma_z^{(i_j)}}{2}, \\
\sigma_z^{(i)} &\mapsto S_p(i,j)S_p(i,k) \ \frac{\sigma_z^{(i_j)} + \sigma_z^{(i_k)}}{2}, \\
I^{(i)} &\mapsto \frac{1 + \sigma_z^{(i_j)}\sigma_z^{(i_k)}}{2}.
\end{align}

Here:
\begin{itemize}
    \item $\sigma_z^{(i_j)}$ acts on copy $j$ of logical qubit $i$.
    \item $S_p(i,j)S_p(i,k)$ enforces the correct relative phase for sector $p$.
\end{itemize}

\subsection{Derivation of the Mapping Rules}

The mapping is constructed so that:
\begin{enumerate}
    \item All $X$ and $Y$ operators are converted to bilinear $Z$-couplings between copies.
    \item The computational basis states in the $rm$-qubit space correspond to classical spin configurations suitable for Ising solvers.
    \item Sector signs $S_p$ encode symmetry constraints that ensure the original spectrum is preserved within some sector.
\end{enumerate}

We illustrate with $\sigma_x^{(i)}$. The projector identities:
\begin{equation}
|0\rangle\langle 1| + |1\rangle\langle 0| = X, \quad 
|0\rangle\langle 0| = \frac{I+Z}{2}, \quad
|1\rangle\langle 1| = \frac{I-Z}{2},
\end{equation}
allow us to express $X$ as a flip between basis states. In the replicated space, flipping is implemented as a mismatch between $\sigma_z^{(i_j)}$ and $\sigma_z^{(i_k)}$. The factor
\[
\frac{1 - \sigma_z^{(i_j)}\sigma_z^{(i_k)}}{2}
\]
is $1$ when $(i_j)$ and $(i_k)$ differ, $0$ when they are equal, reproducing the flip condition. The sector sign $S_p(i,j)S_p(i,k)$ accounts for phase conventions.

A similar argument with $Y = i(|1\rangle\langle 0| - |0\rangle\langle 1|)$ yields the $\sigma_y$ rule, with an extra imaginary factor $i$. For $\sigma_z$, the average $(\sigma_z^{(i_j)}+\sigma_z^{(i_k)})/2$ encodes aligned measurement in both copies.

\subsection{Example: Mapping a Multi-Pauli Term}

Consider $r=2$ and the BK term $X_2 Y_0 Z_1$. Choose sector $p$ with:
\[
S_p(0,2)=-1, \quad S_p(1,2)=+1, \quad S_p(2,2)=-1.
\]
Then:
\begin{align}
X_2 &\mapsto (-1)(+1) \frac{1 - Z^{(2_1)}Z^{(2_2)}}{2}, \\
Y_0 &\mapsto i\,(+1)(-1) \frac{Z^{(0_2)} - Z^{(0_1)}}{2}, \\
Z_1 &\mapsto (+1)(+1) \frac{Z^{(1_1)} + Z^{(1_2)}}{2}.
\end{align}
Multiplying these replacements yields a $Z$-only expression for the term.

\subsection{Sector Hamiltonians}

For fixed $p$, replacing each factor in each term of $H$ produces:
\begin{equation}
H_p = \sum_{\alpha} c_\alpha \; \mathcal{M}_p \!\left( \bigotimes_{\ell} \sigma_{\alpha_\ell}^{(\ell)} \right),
\end{equation}
where $\mathcal{M}_p$ denotes the sector-specific mapping.

\subsection{Normalization Operator}

In the enlarged Hilbert space, a single logical state $|\phi_i\rangle$ may appear multiple times. The normalization operator counts these occurrences:
\begin{equation}
C_p = \sum_{\{\pm\}} \sum_i b_i \; |\phi_i\rangle\langle\phi_i|,
\end{equation}
where $\{\pm\}$ runs over all configurations consistent with $p$ and $b_i$ is the multiplicity.

\subsection{Ground-State Search}

The sector problem is cast as:
\begin{equation}
D_{p,\lambda} = H_p - \lambda C_p.
\end{equation}
By tuning $\lambda$ until the lowest eigenvalue of $D_{p,\lambda}$ approaches zero from below, we obtain $\lambda'_p$, the ground-state energy estimate in sector $p$. The overall ground-state energy is:
\[
\lambda' = \min_p \lambda'_p.
\]

\subsection{Physical Interpretation}

The XBK method encodes logical qubit operators in terms of Ising couplings between a reference copy and its replicas, with sectors enumerating possible alignment patterns. The embedding enlarges the search space so that an Ising solver can reproduce the original quantum ground state without using $X$ or $Y$ operators. The price is an increase from $m$ to $rm$ qubits before quadratization, but the procedure is fully systematic and symmetry-aware.

\section{ Application to Molecular Systems and GUI Development}
\label{sec:appendix_benzene_gui}

To demonstrate the practical utility of the pipeline on complex chemical systems, we performed a series of ground state energy calculations on the benzene molecule. This serves as a benchmark for moderately large, polyatomic systems where electronic correlation and geometric structure are intrinsically linked.


\subsection{Benzene: Multi-parameter Ground State Optimization}

We performed a detailed study of the ground state energy surface for benzene using Dwave DIMOD (C$_6$H$_6$), focusing on the interplay between C--C and C--H bond length variations. These calculations provide insight into how structural optimization in aromatic systems is reflected in the underlying quantum energy landscape.

\begin{figure}[H]
    \centering
    \includegraphics[width=0.4\textwidth]{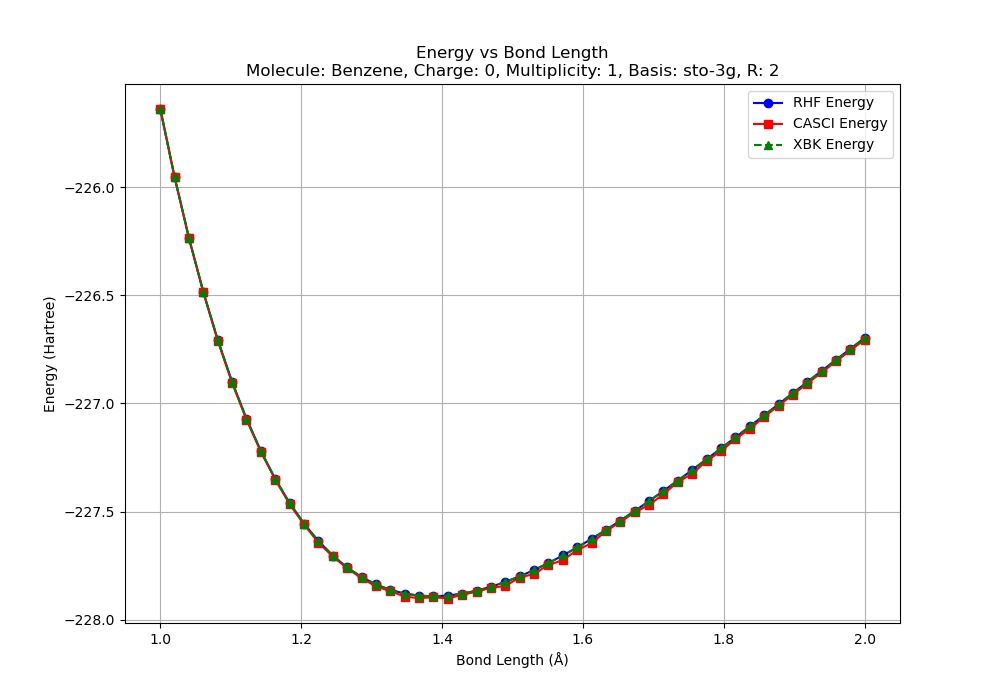}
    \caption{\justifying
        XBK ground state energy for benzene as a function of C--C bond length, with separate curves for fixed C--H bond distances. The global minimum, annotated, corresponds closely to the equilibrium geometry of benzene: lowest energy at C--H = 1.09~\AA, C--C = 1.3684~\AA.
    }
    \label{fig:benzene_multi}
\end{figure}

Figure~\ref{fig:benzene_multi} presents the results of a two-parameter energy scan, where the ground state energy was optimized over a grid of C--C and C--H bond lengths using the XBK encoding. Each curve represents a constant C--H bond length. The plot reveals that the minimum energy is achieved when the C--H bond is set near its natural (experimental) bond length in benzene, demonstrating that quantum-annealing-ready encodings can faithfully resolve structural optima in polyatomic systems.

\subsubsection*{Single-parameter C--C Bond Scan}
\begin{figure}[H]
    \centering
    \includegraphics[width=0.4\textwidth]{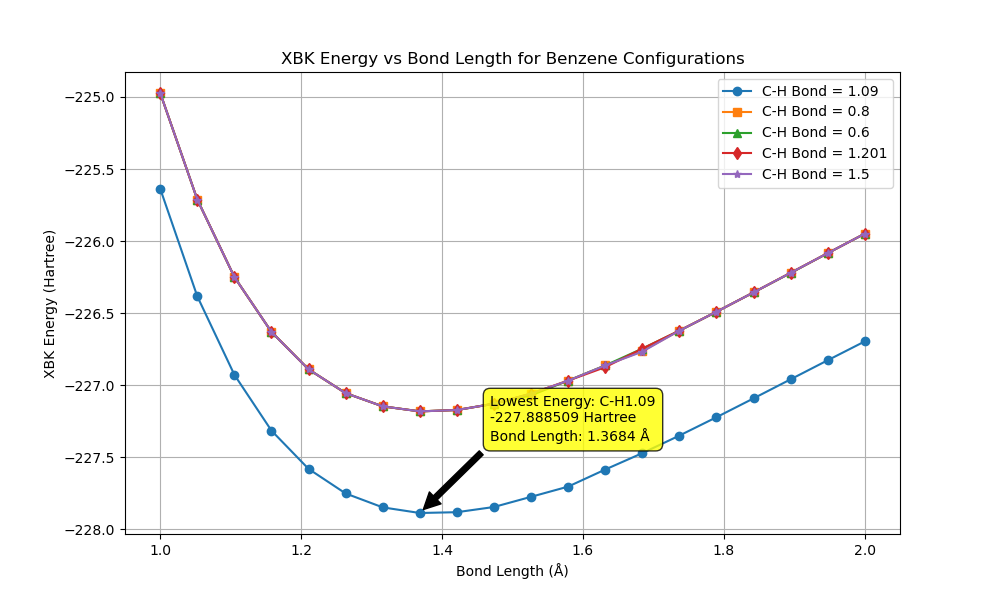}
    \caption{
        DWave DIMOD ground state energies for benzene as a function of C--C bond length with C--H bond fixed at 1.09~\AA. Energies from RHF, CASCI, and XBK encoding are shown and are in excellent agreement; the minimum aligns with the known equilibrium geometry.
    }
    \label{fig:benzene_ccsingle}
\end{figure}

To complement the above, we also performed a single-parameter optimization of the benzene ground state, varying only the C--C bond length while keeping the C--H distance fixed at the equilibrium value. As shown in Figure~\ref{fig:benzene_ccsingle}, the minimum of the energy curve aligns with the expected C--C spacing, consistent with both XBK and conventional quantum chemistry methods.

\subsubsection*{Discussion}
These simulations highlight the accuracy and molecular detail achievable with symmetry-adapted quantum encodings, even for a moderately large system like benzene. The global minimum of the multi-bond surface coincides with the experimentally observed structure, underscoring the method’s ability to capture energetic preferences.

\section{GUI Development}
For the ease of use purpose we are currently developing a GUI based interface, which after its full implementation will be made publically available. 
Here, we are attaching some of the screenshots of the current and previous versions.

\begin{widetext}
\noindent
\begin{minipage}[t]{0.48\columnwidth}
    \begin{figure}[H]
        \centering
        \begin{subfigure}[b]{\textwidth}
            \centering
            \includegraphics[width=0.85\textwidth]{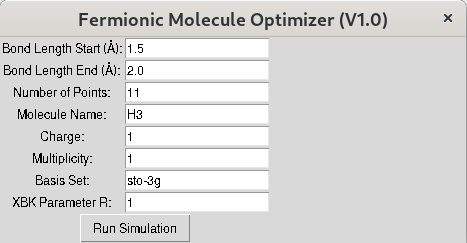}
            \caption{Initial version (v1.0)}
            \label{fig:v1.0}
        \end{subfigure}
        \vspace{2mm}
        \begin{subfigure}[b]{\textwidth}
            \centering
            \includegraphics[width=0.85\textwidth]{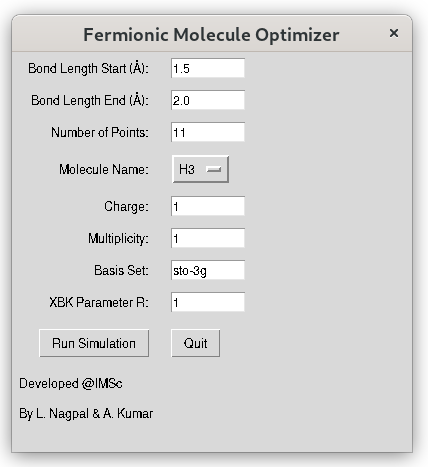}
            \caption{\justifying Updated version (v1.1) with added functionality}
            \label{fig:v1.1}
        \end{subfigure}
        \caption{\justifying Comparison of initial (v1.0) and updated (v1.1) GUI versions.}
        \label{fig:gui_versions_comparison}
    \end{figure}
\end{minipage}
\hfill
\begin{minipage}[t]{0.48\columnwidth}
    \begin{figure}[H]
        \centering
        \includegraphics[width=0.95\textwidth]{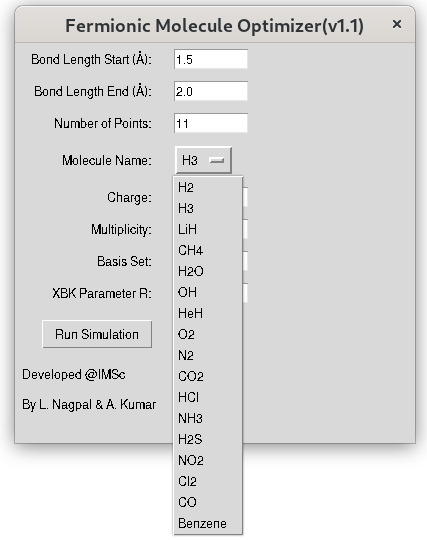}
        \caption{Drop Down list of Molecules in v1.1}
        \label{fig:drop}
    \end{figure}
\end{minipage}
\end{widetext}

\end{document}